\documentclass[12pt]{article}
\usepackage[letterpaper, margin=1.25in]{geometry}
\usepackage{amsmath,amssymb,amsthm,graphicx}
\usepackage{enumitem}
\usepackage{hyperref}
\usepackage{setspace}
\usepackage{indentfirst}

\allowdisplaybreaks

\newcommand{\E}{\mathbf{E}}
\newcommand{\Var}{\mathrm{Var}}

\newcommand{\independent}{\perp \!\!\! \perp}

\newtheorem{theorem}{Theorem}

\newtheorem{proposition}{Proposition}
\newtheorem{lemma}{Lemma}
\newtheorem{assumption}{Assumption}

\newtheorem{example}{Example}

\newlist{assenumerate}{enumerate}{1}  
\setlist[assenumerate]{label=\alph*., font=\bfseries}

\usepackage{tikz}
\usepackage{natbib,har2nat}

\title{\Large Distributional treatment effect with latent rank invariance\thanks{This work was supported by the European Research Council through the grant ERC-2018-CoG-819086-PANEDA, while the author was employed as a postdoctoral researcher at the University of Oxford. The author declares no conflict of interest.}}
\date{\large September 13, 2026 
}
\author{\large Myungkou Shin\thanks{School of Social Sciences, University of Surrey. email: \href{mailto:m.shin@surrey.ac.uk}{m.shin@surrey.ac.uk}.}}

\begin{document}

	\maketitle

	\begin{abstract}
		Treatment effect heterogeneity is of great concern in policy evaluation. However, existing analyses have mostly been limited to summary measures such as an average treatment effect, due to the fundamental limitation that we cannot simultaneously observe both potential outcomes for a given unit. In this paper, I identify and estimate moment-identified distributional treatment effect (DTE) parameters, such as the marginal distribution of treatment effect. The key identifying assumption is that the two potential outcomes are conditionally independent given a latent variable. Interpreting this latent variable as underlying individual-level heterogeneity, I motivate the identifying assumption as ‘latent rank invariance.' In the conditional independence framework, the DTE parameters are identified given two proxy variables. In implementation, I assume a finite support on the latent variable, imposing a finite mixture structure on the identifying assumption. Using Neyman orthogonality, I establish asymptotic normality of the estimator, enabling inference for DTE parameters.
	\end{abstract}
	
	\noindent \hspace{5mm} \textbf{Keywords}:  distributional treatment effect, proximal inference, finite mixture, \\
	\text{ } \hspace{25.8mm} $U$-statistic, Neyman orthogonality.
	
	\pagebreak 

	\section{Introduction}\label{sec:intro}
	
	The fundamental limitation that we cannot simultaneously observe the two potential outcomes\textemdash treated potential outcome and untreated potential outcome\textemdash for a given unit makes the task of identifying the distribution of treatment effect particularly challenging. Thus, researchers often focus on some summary measures of the treatment effect distribution, such as the average treatment effect (ATE) or the quantile treatment effect (QTE), and abstract from the full picture of treatment effect. 
	
	However, there still remain a lot of questions that can be answered only with the \textit{distribution} of the treatment effect. Consider a potential outcome setup with a binary treatment: $$
	Y= D \cdot Y(1) + (1-D) \cdot Y(0).
	$$ 
	$Y(1)$ is a treated potential outcome, $Y(0)$ is an untreated potential outcome, and $D \in \{0,1\}$ is a binary treatment variable. Many policy-relevent questions directly relate to the distributional aspect of the treatment effect: is the treatment Pareto improving?; how heterogeneous is the treatment effect at the unit level?; how many people would select into treatment when the cost of opting in is $c$? These questions correspond to testing $H_0: \Pr \{ Y(1)-Y(0) \leq 0\}=0$ and estimating $\Var \big( Y(1)-Y(0) \big)$ and $\Pr \{Y(1)-Y(0) \geq c\}$. Note that these quantities, $\Pr \{ Y(1)-Y(0) \leq 0\}, \Pr \{Y(1)-Y(0) \geq c\}$ and $\Var \big( Y(1)-Y(0) \big)$, all come from the distribution of individual-level treatment effect $Y(1) - Y(0)$. 
	
	The goal of this paper is to answer these policy questions with distributional concerns. Specifically, I focus on a class of distributional treatment effect (DTE)\footnote{Some previous works in the literature use the terminology `distributional effect' to discuss parameters that are a functional of the two marginal distributions of the potential outcomes; e.g., \citet{FP2016}. To avoid confusion, I reserve the expression `distributional' for only when the object involves the joint distribution of the two potential outcomes.} parameters identified by a moment function defined over the two potential outcomes: \begin{align}
	\E \left[ m \big( Y(1), Y(0) ; \theta \big) \right] = 0. \label{eq:original_moment_intro}
	\end{align}
	The DTE parameter $\theta$ is identified if the joint distribution of the two potential outcomes is identified. Examples of $\theta$ include distribution of treatment effect, variance of treatment effect, and many more.
	
	When we believe that there is no dependence between the two potential outcomes, meaning that a realized value of the treated potential outcome has no information on the individual-level heterogeneity and thus has no predictive power for the untreated potential outcome and vice versa, identification of the joint distribution of the two potential outcomes becomes trivial with a randomized treatment. Once we identify the marginal distributions of the two potential outcomes, the joint distribution becomes their product. However, this assumption is extremely restrictive. Thus, I instead impose \textit{conditional} independence, as in \citet{CHH}, assuming that there exists a latent variable which captures the dependence between the two potential outcomes. 

	Consider a simple additive model: the two potential outcomes are constructed with a individual-level latent variable $U \in \mathcal{U}\subset \mathbb{R}$ and two regime-specific random shocks $\varepsilon^1$ and $\varepsilon^0$: \begin{align}
	Y(1) &= \mu^1(U) + \varepsilon^1, \label{eq:additive1} \\
	Y(0) &= \mu^0(U) + \varepsilon^0. \label{eq:additive2}
	\end{align}
	When the noises are truly random, satisfying $\varepsilon(1) \independent \varepsilon(0) \text{ } \big| \text{ } U$, we can characterize the joint distribution of the two potential outcome as follows: $$
	\Pr \left\lbrace Y(1) \leq y_1, Y(0) \leq y_0 \right\rbrace = \E \left[ \Pr \left\lbrace Y(1) \leq y_1 | U \right\rbrace \cdot \Pr \left\lbrace Y(0) \leq y_0 |U \right\rbrace \right]. 
	$$
	Thus, the task of identifying the joint distribution of the two potential outcomes becomes that of identifying the conditional distribution of $Y(1)$ given $U$, the conditional distribution of $Y(0)$ given $U$, and the marginal distribution of $U$. 
	
	To identify the conditional distribution of $Y(d)$ given $U$ and the marginal distribution of $U$, I assume that there are two additional proxy variables $X,Z$ that are conditionally independent of each other and the potential outcomes, given $U$. This identification strategy is drawn from the nonclassical measurement error literature and the proximal inference literature: see \citet{HS2008,MGT,D2023,K2023,N2022} and more. In the simple example \eqref{eq:additive1}-\eqref{eq:additive2}, the proxy variables $X,Z$ will shift $\mu^d(U)$ independently of $\big( \varepsilon^1, \varepsilon^0 \big)$, allowing us to decompose the variation of $Y(d)$ into the variation of $U$ and the variation of $\varepsilon^d$. Additionally, to find a labeling on $U$, I assume that the conditional distribution of $Y(d)$ given $U$ orders $U$: latent rank invariance.

	In implementation, I propose a two-step estimation strategy to estimate the DTE parameters, assuming a finite support on $U$.\footnote{Though I assume that $U$ is finitely discrete in the estimation, the identification result does not require such a restriction and I develop an alternative estimation method based on sieve maximum likelihood for a setup with continuous $U$, in the Supplemental Appendix.} Under the finite support assumption, the conditional independence framework becomes a finite mixture. In the first step, I propose a novel nonlinear least-square estimator to estimate the nonparametric finite mixture. In the second step, I show that the infeasible moment function in \eqref{eq:original_moment_intro} can be modified to a feasible moment function, with the finite mixture parameters as nuisance parameters. Asymptotic normality of the two-step plug-in DTE estimator is established, taking advantage of the Neyman orthogonality. 

	As an empirical illustration, I revisit \citet{JMR} and estimate the effect of a workplace wellness program eligibility on employees' medical spending. Using the DTE framework, I explore the treatment effect beyond the original paper's scope, estimating the entire distribution of the treatment effect. The DTE estimate demonstrates clear information gain compared to partial bounds.

	This paper makes a contribution to the distributional treatment effect literature: see \citet{BBDM} for an overview. As a working example of a DTE parameter, I point identify and estimate the marginal distribution of treatment effect. This provides us much information gain compared to the existing partial identification results in the literature: \citet{FP2010,FSS2014,FR2019,FL2021,KC2023} and more. A few papers that have instead focused on point identification results are \citet{HSC,CHH}. The closest to this paper is \citet{CHH}; this paper follows their conditional independence approach and contributes by proposing a flexible estimation strategy which does not rely on parametric distributions. Other works that discuss estimation of DTE are \citet{N2023,ahn2025prognostic}. Compared to this paper, they rely on either unconditional independence, or conditional independence with an observed conditioning variable, both of which are arguably more restrictive than this paper's conditional independence framework with a latent conditioning variable. 

	This paper also contributes to the nonclassical measurement error/proximal inference literature and the nonparametric finite mixture literature: \citet{H2008,HS2008,HKS2014,BJR,D2023,N2022} and more. Unlike existing estimators based on the diagonalization, the nonlinear least-square estimator in this paper explicitly impose nonnegativity constraints on probabilities. Monte Carlo simulations show significant gain in the finite-sample performance, both on the extensive and intensive margins; notably, the new estimator is less susceptible to numerical singularity and consequently has higher estimation completion rate. In addition, while establishing asymptotic normality of the DTE estimator, I develop an orthogonalization procedure for inference on moment-based parameters in a nonparametric finite mixture model, which has broader applicablity beyond the DTE estimation in this paper.

	The rest of the paper is organized as follows. Section \ref{sec:model} describes the conditional independence framework of this paper and discusses DTE parameters identified by a moment condition \eqref{eq:original_moment_intro}. Section \ref{sec:id} identifies the joint distribution of the two potential outcomes and thus the DTE parameters. Section \ref{sec:est} explains the estimation method for the DTE parameters and develops asymptotic theory for the estimator. Section \ref{sec:sim} contains Monte Carlo simulation results and Section \ref{sec:emp} applies the estimation procedure to an empirical dataset from \citet{JMR}.  
	
	\section{Model}\label{sec:model}

	An econometrician observes a dataset $\left\lbrace Y_{i}, D_i, X_i, Z_i \right\rbrace_{i=1}^n$ where $Y_{i}, X_i, Z_i \in \mathbb{R}$ and $D_i \in \{0,1\}$. $Y_i$ is an outcome of interest and $D_i$ is a binary treatment status variable. $Y_i$ comes from two potential outcomes: \begin{align}
		Y_{i} = D_{i} \cdot Y_{i}(1) + \left(1 - D_{i}\right) \cdot Y_{i}(0). \label{eq:potential_outcome}
	\end{align}
	In addition to $\left(Y_i(1), Y_i(0), D_i, X_i, Z_i \right)$, there is a latent variable $U_i \in \mathcal{U} \subset \mathbb{R}$ that models the individual-level heterogeneity. $U_i$ plays a key role in putting restrictions on the joint distribution of $Y_i(1)$ and $Y_i(0)$ and overcoming the fundamental limitation that we observe only one potential outcome for a given unit. Since $U_i$ is latent, I assume that there exist two proxy variables $X_i$ and $Z_i$ that are informative for $U_i$. The dataset comes from random sampling: $\left( Y_i(1), Y_i(0), D_i, X_i, Z_i, U_i \right) \overset{iid}{\sim} \mathcal{F}$. 
	
	Firstly, I assume conditional random assignment on the treatment $D_i$ and exclusion restriction on the proxy variable $Z_i$.
	
	\begin{assumption}\label{ass:assignment} \textit{(assignment/exclusion restriction)} \[
	\big(Y_i(1), Y_i(0), X_i \big) \independent \big( D_i, Z_i \big) \text{ } | \text{ } U_i.	
	\]
	\end{assumption}
	
	\noindent Assumption \ref{ass:assignment} assumes that the treatment is as good as random with regard to the potential outcomes and $X_i$ after conditioning on the latent variable $U_i$. In this sense, Assumption \ref{ass:assignment} is a restriction on treatment endogeneity. In addition, Assumption \ref{ass:assignment} assumes that the proxy variable $Z_i$ does not have any additional information on the potential outcomes after conditioning on the latent variable $U_i$, satisfying exclusion restriction. Note that Assumption \ref{ass:assignment} does not impose any restriction on the dependence between $Z_i$ and $D_i$. The proxy variable $Z_i$ may still depend on treatment. This is a standard assumption in the proximal inference literature, making $X_i$ the outcome-aligned proxy and $Z_i$ the treatment-aligned proxy: see \citet{D2023,N2022,MGT} for more.
	
	For the rest of the paper, I focus on randomly assigned treatments, focusing on randomized controlled trials.
	
	\vspace{4mm} 

	\noindent \textit{Remark 1.} \textit{A sufficient condition for Assumption \ref{ass:assignment} is } \begin{align*}
	\big( Y_i(1), Y_i(0), X_i, U_i \big) \independent D_i \text{ and } \big( Y_i(1), Y_i(0), X_i \big) \independent Z_i \text{ } | \text{ } ( D_i, U_i). 
	\end{align*}

	\noindent For example, a control covariate measured at baseline can be used as the proxy variable $X_i$. If another variable $Z_i$ is measured at follow-up and only depends on the pretreatment unobserved heterogeneity $U_i$ and treatment $D_i$, it can be used as the second proxy variable. 
	
	When $U_i$ is observed, Assumption \ref{ass:assignment} identifies treatment effect parameters such as average treatment effect (ATE) and quantile treatment effect (QTE). However, even when $U_i$ is observed, we still cannot identify the distribution of treatment effect since Assumption \ref{ass:assignment} does not tell us anything about the dependence between $Y_i(1)$ and $Y_i(0)$. 
	
	To impose restrictions on the joint distribution of $Y_i(1)$ and $Y_i(0)$, I assume that the latent variable $U_i$ captures all of the dependence between the two potential outcomes and the proxy variable $X_i$. 
	
	\begin{assumption}\label{ass:condind}
		\textit{(conditional independence)} $Y_i(1), Y_i(0)$ and $X_i$ are all mutually independent given $U_i$. 
	\end{assumption}
	
	\noindent Two parts of Assumption \ref{ass:condind} serve different purposes. Firstly, the conditional independence between $(Y_i(1), Y_i(0))$ and $X_i$ assumes that the proxy variable $X_i$ does not give us additional information for the outcome variable given $U_i$. This is a standard assumption in the nonclassical measurement error literature: \citet{H2008,HS2008}. Additionally, Assumption \ref{ass:condind} assumes that the two potential outcomes are independent of each other given $U_i$. This is the key assumption that identifies the joint distribution of the two potential outcomes. 

	When $U_i$ is observed, Assumptions \ref{ass:assignment}-\ref{ass:condind} identify the joint distribution of the two potential outcomes and various DTE parameters, such as distribution of treatment effect. Since $U_i$ is not observed, identifying the conditional densities of $Y_i(1), Y_i(0)$ given $U_i$ and the marginal density of $U_i$ remains the main challenge for identification. 

	To motivate Assumption \ref{ass:condind}, I present two examples based on widely adopted econometric models. The first example is a repeated measurement model: \citet{CHH,CH2008,CHS,ACFMR} and more. 

	\begin{example}\label{eg:measure} (repeated measures) The econometrician observes an outcome of interest $Y_i$, a binary treatment $D_i$, and two proxy variables $X_i$ and $Z_i$ which measure an innate ability $U_i$. $Y_i = D_i \cdot Y_i(1) + (1-D_i) \cdot Y_i(0)$ and \begin{align*}
		Y_i(d) &= \mu^{d} + \alpha^{d} U_i + \varepsilon_{i}^{d} \hspace{4mm} \text{for } d=0,1, \\
		X_i &= \mu^{X} + \alpha^{X} U_i + \varepsilon_{i}^{X}, \\
		Z_i &= \mu^{Z} + \alpha^{Z} U_i + \varepsilon_{i}^{Z}.
	\end{align*}
	$\varepsilon_i^{0}, \varepsilon_i^{1}, \varepsilon_i^X$ and $\varepsilon_i^Z$ are mutually independent given $U_i$. When treatment $D_i$ is assigned randomly, Assumptions \ref{ass:assignment}-\ref{ass:condind} are satisfied. 
	\end{example}

	\noindent This example is from \citet{ACFMR}. \citet{ACFMR} studies the effect of early childhood intervention on cognitive and socio-emotional skills of children aged 12 to 24 months old. The randomized intervention included home visits that provided parenting advice to parents. In this setup, the cognitive ability score measured at follow-up is the outcome $Y_i$ and the maternally reported baseline measures of cognitive ability such as the number of words the child can say and the number of complex phrases the child can say are the proxy variables $X_i$ and $Z_i$. Then, the child's latent cognitive ability at baseline would be the latent variable $U_i$.\footnote{Since \citet{ACFMR}'s analysis discusses multiple measures of cognitive and socio-emotional abilities, other variables can similarly be used as $(Y_i, X_i, Z_i)$, with a possibly different interpretation on $U_i$.} The only additional assumption that I impose in Example 1 compared to \citet{ACFMR}'s model is that measurement errors $\varepsilon_i^0$ and $\varepsilon_i^1$ are conditionally independent of each other given the innate ability $U_i$. 

	The second example is the hidden Markov model: \citet{KS2009,AM2011,HS2012,HY2018} and more.
	\begin{example}\label{eg:panel} (hidden Markov model) The econometrician observes $\left\lbrace \{Y_{it} \}_{t=1}^3, D_i \right\rbrace_{i=1}^n$ where $
		Y_{it}(d) = g_{d} \big( V_{it}, \varepsilon_{it}^d \big)
		$ for $t=1,2,3$ and $d=0,1$ and \begin{align}
		Y_{it} &= \begin{cases} Y_{i1}(0) & \text{ if } t=1 \\
			D_i \cdot Y_{it}(1) + (1 - D_i) \cdot Y_{it}(0) & \text{ if } t=2, 3 \end{cases}. \notag 
	\end{align}
	The latent state process $\left\lbrace V_{it} \right\rbrace_{t=1}^3$ is first-order Markovian given $D_i$. Also, the time-by-treatment-status shocks $ \varepsilon_{i1}^0, \varepsilon_{i2}^0, \varepsilon_{i2}^1, \varepsilon_{i3}^0, \varepsilon_{i3}^1$ and $\left( \left\lbrace V_{it} \right\rbrace_{t=1}^3, D_i \right)$ are all mutually independent. When treatment $D_{i}$ is randomly assigned at time $t=2$, i.e. $\{ V_{it} \}_{t=1}^2 \independent D_i$, Assumptions \ref{ass:assignment}-\ref{ass:condind} are satisfied with $Y_i = Y_{i2}, X_i = Y_{i1}, Z_i = Y_{i3}$ and $U_i = V_{i2}$. 
	\end{example}

	\noindent Example \ref{eg:panel} is an extension of the hidden Markov model to a potential outcome setup following \citet{D2023}. The additional assumption I make here is that the idiosyncratic error $\varepsilon_{it}^d$ are independent across treatment status. To provide a context, let us connect this example to \citet{JMR} from Section \ref{sec:emp}. In \citet{JMR}, the authors randomly assigned workplace wellness program eligibility to university employees and estimated its effect on individual medical spending. The medical spending information was collected before, during, and after the treatment. Applying Example \ref{eg:panel} to their empirical context, the pretreatment and the follow-up medical spending are the proxy variables $X_i$ and $Z_i$ and the concurrent medical spending is the outcome variable $Y_i$. The shared latent state $V_{it}$ adopts the natural interpretation As underlying health status.

	The common feature shared across the two examples is that the treatment $D_i$ affects the outcome $Y_i$ in a regime-changing manner; there are two separate production functions\textemdash $\mu^1 + \alpha^1 U_i + \varepsilon_i^1$ and $\mu^0 + \alpha^0 U_i + \varepsilon_i^0$ in the first example, and $g_0(V_{i2}, \varepsilon_{i2}^0)$ and $g_1(V_{i2}, \varepsilon_{i2}^1)$ in the second\textemdash and the treatment $D_i$ decides which production function is applied to generate $Y_i$. This point is briefly addressed in \citet{ACFMR} where the authors discuss two possible mechanisms of treatment effect: a change in production function itself and a change in the amount of inputs.\footnote{
		This relates to the two different independence assumptions in the literature: $Y_i(1) \independent Y_i(0)$ and $Y_i(0) \independent \big( Y_i(1) - Y_i(0) \big)$. The latter approach would be plausible if the treatment mechanism works in an ``input-changing'' manner. That is, a treated potential outcome goes through the same outcome generating process as an untreated potential outcome, but with an additional source of heterogeneity, which is independent of the existing source of heterogeneity. Suppose $V_i \independent \varepsilon_i \text{ } \big| \text{ } U_i$ and $$
		Y_i(d) = \alpha + \mu^0 U_i + d \cdot \mu^1 V_i + \varepsilon_i.
		$$
		Then, $Y_i(0) \independent \big( Y_i(1) - Y_i(0) \big) \text{ } \big| \text{ } U_i$. Here, $V_i$ denotes the new source of individual-level heterogeneity, which contributes to the outcome only when treated. An example of such an empirical context is where the treatment provides a new infrastructure, such as electronic devices in education context, etc. 
	} 
	Both this paper and \citet{ACFMR} assume the former, the regime-changing treatment effect meachanism. However, \citet{ACFMR} abstracts away from a comprehensive counterfactual analysis  and does not take a stance on how the error terms in the two production functions would be connected. In this paper, in contrast, assumes that the two error terms are indeed pure random noises and independent of each other conditional on $U_i$. 

	In this sense, the conditional independence framework of this paper is best suited for empirical contexts where a randomized treatment affects an outcome by placing treated units in an alternative outcome generating process. When the parenting guidance systemically changed children's cognitive ability development process, \citet{ACFMR} would fit this framework. Similarly, if the wellness program systemically changed participants' health-related behaviors, \citet{JMR} would also fit the framework. Other examples include a job training program or an educational intervention based on teaching methodology.\footnote{For example, when treated participants are assigned a new job, such as in the National Supported Work Demonstration, the mechanism of how innate aptitude and skill lead to wage income will be shifted. Similarly, by being taught with a different teaching methodology, the production function of students' outcome shifts to a new regime: e.g., \citet{banerjee2007remedying,muralidharan2019disrupting} and more.}

	We can relax Assumptions \ref{ass:assignment}-\ref{ass:condind} in several directions. Firstly, we may extend the conditional independence framework to include observable conditioning variables. Given some observable control covariates $C_i$, Assumptions \ref{ass:assignment}-\ref{ass:condind} under random treatment can be replaced by \begin{gather*}
	\big( Y_i(1), Y_i(0), X_i, C_i, U_i \big) \independent D_i \text{ and } \big( Y_i(1), Y_i(0), X_i \big) \independent Z_i \text{ } | \text{ } ( D_i, C_i, U_i) \\
	Y_i(1), Y_i(0) \text{ and } X_i \text{ are all mutually independent given } \big( C_i, U_i \big).
	\end{gather*}
	In this setup, the latent variable $U_i$ only needs to explain remaining heterogeneity among units that is not explained by $C_i$. Secondly, we can model the latent variable $U_i$ to be multidimensional, as long as the two proxy variables $X_i, Z_i$ are of the same dimension. The specific treatment of multidimensional $U_i$ is provided in the Supplemental Appendix.

	\subsection{Distributional treatment effect parameters}\label{subsec:DTE}

	The goal of this paper is to identify and estimate distributional treatment effect (DTE) parameters. In particular, I focus on moment-identified DTE parameters: \[
	\E \left[ m \big( Y_i(1), Y_i(0) ; \theta \big)\right] = 0. 
	\]
	Below I provide several examples of moment-identified DTE parameter $\theta$: with $F_{Y(1)-Y(0)}$ denoting the marginal distribution of $Y_i(1)-Y_i(0)$, \begin{itemize}
	\item[-] Conditional ATE given $Y_i(0)$: $\theta(\mathcal{Y}) = \E[ Y_i(1) - Y_i(0) | Y_i(0) \in \mathcal{Y}]$.
	\item[-] Variance-adjusted ATE: $\theta(\zeta)= \E [Y_i(1) - Y_i(0)] + \dfrac{\zeta}{2} \Var \big( Y_i(1) - Y_i(0) \big)$.
	\item[-] Share of winners: $\theta = 1 - F_{Y(1)-Y(0)}(0)$.
	\item[-] $\alpha$-th quantile of treatment effect: $\theta(\alpha) = F_{Y(1)-Y(0)}^{-1}(\alpha)$. 
	\end{itemize} 

	The conditional ATE discusses treatment effect heterogeneity in terms of the untreated potential outcome. By conditioning on the untreated outcome, we can determine whether the overall treatment effect is driven by gains on units with low baseline outcomes or those with high baseline outcomes.  

	The variance-adjusted ATE aggregates the treatment effect while allowing for inequality aversion. The risk aversion in expected utility maximization directly translates to the inequality aversion in social welfare. Thus, the quadratic approximation in utility maximization, e.g., \citep{levy1979approximating}, applies to the social welfare context as well: see \citet{epstein1992quadratic} for more. In the approximation, $\zeta \geq 0$ is the inequality aversion parameter.
	
	The share of winners measures the proportion of units that are better off under the treatment. When the share of winners is one, the treatment is Pareto-improving, benefiting a treated unit with probability one. When the share of winners is bigger than 0.5, the treatment benefits a treated unit more often than not . 
	
	On the flip side of the distribution function, the $\alpha$-th quantile of treatment effect marks the $100 \cdot \alpha$\% worst-case (or best-case when lower $Y_i$ is more desirable) threshold in treatment effect distribution. This parameter gives us a probabilistic guarantee that the probability of the treatment harming a unit by a margin greater than this threshold does not exceed $\alpha$. Importantly, this parameter is different from the quantile treatment effect at $\alpha$-th quantile, $F_{Y(1)}^{-1}(\alpha) - F_{Y(0)}^{-1}(\alpha)$, which does not give us any probabilistic guarantee. 

	These DTE parameters answer important policy questions regarding fairness and treatment effect heterogeneity. However, the existing literature mostly attempts to answer these questions using a partial identification approach, which often provide little information due to the identified set being too large. In this paper, using the conditional independence framework, I point identify and estimate the DTE parameters, giving us significant information gain. 

	\section{Identification}\label{sec:id}

	To identify the DTE parameters under the conditional independence framework, I show that the joint distribution of the two potential outcomes $Y_i(1)$ and $Y_i(0)$ is identified. Fix some $\mathcal{Y}, \mathcal{X}, \mathcal{Z} \subset \mathbb{R}$ and $d \in \{0,1\}$. Under Assumptions \ref{ass:assignment}-\ref{ass:condind}, \begin{align}
	&\Pr \left\lbrace Y_i \in \mathcal{Y} , X_i \in \mathcal{X} \, | \, D_i=d, Z_i \in \mathcal{Z} \right\rbrace \notag \\
	&= \int_{\mathcal{U}} \Pr \left\lbrace Y_i(d) \in \mathcal{Y} \, | \, U_i = u \right\rbrace \Pr \left\lbrace X_i \in \mathcal{X} \, | \, U_i = u \right\rbrace f_{U|D,Z}(u \, | \, D_i=d, Z_i \in \mathcal{Z} ) du.\label{eq:finite_mixture}
	\end{align}
	For each subpopulation $\left\lbrace i: (D_i, Z_i) \in \{d\} \times \mathcal{Z} \right\rbrace$, the conditional density of $(Y_i, X_i)$ admits a mixture model. Also, the subpopulations share the same mixture component distributions $\Pr \left\lbrace Y_i(d) \in \mathcal{Y} \, | \, U_i = u \right\rbrace \cdot \Pr \left\lbrace X_i \in \mathcal{X} \, | \, U_i = u \right\rbrace$ as long as their treatment status are the same. Since $Y_i(1) \independent Y_i(0) \; | \; U_i$, the identification of the joint distribution of $Y_i(1)$ and $Y_i(0)$ is the identification of this mixture model. 

	To see that the mixture model is identified, I turn to the nonclassical measurement error literature: \citet{H2008,HS2008}. In a setup without a treatment variable $D_i$, \citet{H2008,HS2008} identifies the mixture model using eigenvalue decomposition. Specifically, \citet{H2008} assumes $U_i$ to be finitely discrete and \citet{HS2008} extends this to allow for continuous $U_i$. In this paper, I build on their diagonalization results and develop an identification result for a setup with a binary treatment $D_i$ that affects the mixture component distribution.  
	
	For a finitely discrete $U_i$, I use Assumption \ref{ass:id}. Assumption \ref{ass:id} is a modification of the identifying assumption from \citet{H2008}:
	\begin{assumption}\label{ass:id}
		\text{ }  \begin{assenumerate}
			\item \textit{(finitely discrete $U_i$)} $\mathcal{U} = \{u^1, \cdots, u^K\}$. 

			\item \textit{(full rank)} There exist two partitions on $\mathbb{R}$, $\{\mathcal{X}^m\}_{m=1}^{M_X}$ and $\{\mathcal{Z}^m\}_{m=1}^{M_Z}$,\footnote{For example, with some $\{x^m\}_{m=0}^{M_X}$ such that $-\infty=x^0 < x^1 < \dots < x^{M_X} = \infty$, $\mathcal{X}^m = \left(x^{m-1},x^m \right]$ for $m=1,\dots,M_X$.} 
		such that three matrices {\allowdisplaybreaks
			\begin{align*}
			\Gamma_X &:= \scalebox{0.95}{$\begin{pmatrix} \Pr \left\lbrace X_i \in \mathcal{X}^1 \, | \, U_i = u^1 \right\rbrace & & \Pr \left\lbrace X_i \in \mathcal{X}^1 \, | \, U_i = u^K \right\rbrace \\
			\vdots & \ddots & \vdots \\
			\Pr \left\lbrace X_i \in \mathcal{X}^{M_X} \, | \, U_i = u^1 \right\rbrace & & \Pr \left\lbrace X_i \in \mathcal{X}^{M_X} \, | \, U_i = u^K \right\rbrace \end{pmatrix}$}, \\ 
			\Lambda_0 &:= \scalebox{0.95}{$\begin{pmatrix} \Pr \left\lbrace U_i = u^1 \, | \, D_i = 0, Z_i \in \mathcal{Z}^1 \right\rbrace & & \Pr \left\lbrace U_i = u^1 \, | \, D_i = 0, Z_i \in \mathcal{Z}^{M_Z} \right\rbrace \\
			\vdots & \ddots & \vdots \\
			\Pr \left\lbrace U_i = u^K \, | \, D_i = 0, Z_i \in \mathcal{Z}^1 \right\rbrace & & \Pr \left\lbrace U_i = u^K \, | \, D_i = 0, Z_i \in \mathcal{Z}^{M_Z} \right\rbrace \end{pmatrix}$}, \\
			\Lambda_1 &:= \scalebox{0.95}{$\begin{pmatrix} \Pr \left\lbrace U_i = u^1 \, | \, D_i = 1, Z_i \in \mathcal{Z}^1 \right\rbrace & & \Pr \left\lbrace U_i = u^1 \, | \, D_i = 1, Z_i \in \mathcal{Z}^{M_Z} \right\rbrace \\
			\vdots & \ddots & \vdots \\
			\Pr \left\lbrace U_i = u^K \, | \, D_i = 1, Z_i \in \mathcal{Z}^1 \right\rbrace & & \Pr \left\lbrace U_i = u^K \, | \, D_i = 1, Z_i \in \mathcal{Z}^{M_Z} \right\rbrace \end{pmatrix}$}
			\end{align*}
			}
			have rank $K$.
			
			\item \textit{(no repeated eigenvalue)} For any $k \neq k'$, there exist some $\mathcal{Y}, \mathcal{Y}' \subset \mathbb{R}$ such that \begin{align*}
			\Pr \left\lbrace Y_i(0) \in \mathcal{Y} \, \big| \, U_i = u^k \right\rbrace &\neq \Pr \left\lbrace Y_i(0) \in \mathcal{Y} \, \big| \, U_i = u^{k'} \right\rbrace, \\
			\Pr \left\lbrace Y_i(1) \in \mathcal{Y}' \, \big| \, U_i = u^k \right\rbrace &\neq \Pr \left\lbrace Y_i(1) \in \mathcal{Y}' \, \big| \, U_i = u^{k'} \right\rbrace.
			\end{align*}
		\end{assenumerate}
	\end{assumption}

	\noindent $K$ is the number of points in the support of $U_i$. Assumption \ref{ass:id}.b implicitly assumes that the support of $X_i$ and that of $Z_i$ have at least $K$ points: $M_X, M_Z \geq K$. This restriction is sensible since I use the proxy variables $X_i, Z_i$ to capture the variation in the latent variable $U_i$. Applying the mixture model terminology, Assumption \ref{ass:id}.b assumes that the mixture component distributions of $X_i$ are distinct ($\Gamma_X$ is full rank) and that the proxy variable $Z_i$ creates sufficient variation in mixture weights ($\Lambda_0$, $\Lambda_1$ are full rank). 

	For continuous $U_i$, I draw from \citet{HS2008}. Let $f_{Y(d)|U}$ denote the conditional density of $Y_i(d)$ given $U_i$, $f_{X|U}$ denote the conditional density of $X_i$ given $U_i$, and $f_{U|D=d,Z}$ denote the conditional density of $U_i$ given $D_i=d$ and $Z_i$, for $d=0,1$. Define integral operators $L_{X|U}$ and $L_{Z|D=d,U}$ that map a function in $\mathcal{L}^1(\mathbb{R})$ to a function in $\mathcal{L}^1(\mathbb{R})$: for $d=0,1$, \begin{align*}
		\left[ L_{X|U} g \right] (x) &= \int_{\mathbb{R}} f_{X|U}(x|u) g(u) du, \\
		\left[ L_{Z|D=d,U} g \right] (z) &= \int_{\mathbb{R}} f_{Z|D=d,U}(z|u) g(u) du.
	\end{align*}
	
	\begin{assumption}\label{ass:id_cont}
		Assume \begin{assenumerate}
			\item \textit{(continuous $U_i$)} $\mathcal{U} = [0,1]$. 

			\item \textit{(bounded density)} The conditional densities $f_{Y(1)|U}, f_{Y(0)|U},f_{X|U},f_{U|D=1,Z}$ and $f_{U|D=0,Z}$ and the marginal densities $f_U,f_{Z|D=1}$ and $f_{Z|D=0}$ are bounded. 
			
			\item \textit{(completeness)} The integral operators $L_{X|U}, L_{Z|D=1,U}$ and $L_{Z|D=0,U}$ are injective on $\mathcal{L}^1(\mathbb{R})$.
			
			\item \textit{(no repeated eigenvalue)} For any $u \neq u'$ and $d \in \{0,1\}$, \\ $
			\Pr \left\lbrace f_{Y(d)|U}(Y_i|u) \neq f_{Y(d)|U}(Y_i|u') | D_i=d \right\rbrace> 0.
			$
		\end{assenumerate}
	\end{assumption}

	\noindent Assumption \ref{ass:id_cont} comprises Assumptions 1-4 of \citet{HS2008}, and replaces Assumption \ref{ass:id}.

	When $U_i$ is continuous, we need an additional condition compared to Assumption \ref{ass:id}, to find an ordering on the infinite collection $
	\{f_{Y(1)|U}(\cdot |u), f_{Y(0)|U}(\cdot |u), f_{X|U}(\cdot | u) \}_u.
	$
	The following assumption uses a known functional $M$ to do so.

	\begin{assumption}\label{ass:latent_rank}
		\textit{(latent rank)} There exists a functional $M$ defined on $\mathcal{L}^1(\mathbb{R})$ such that $
		u \mapsto M f_{Y(d)|U} (\cdot| u) 
		$
		defined on $\mathcal{U}$ is strictly increasing and continuously differentiable, for some $d=0,1$.
	\end{assumption}

	\noindent A simple example where Assumption \ref{ass:latent_rank} fails is when $\mathcal{U} = [-1,1]$ and $Y_i(d) \text{ } \big| \text{ } U_i=u \sim \mathcal{N} ( u^2 + d, \sigma^2)$. Neither $f_{Y(1)|U}$ nor $f_{Y(0)|U}$ helps us find an ordering between $f_{X|U} (\cdot | u)$ and $f_{X|U}(\cdot | -u)$. 

	To motivate this, let us revisit the simple example \eqref{eq:additive1}-\eqref{eq:additive2} and suppose that $\E[\varepsilon_i^1 | U_i] =0$ and $\mu^1 (u)$ is strictly increasing in $u$. Then, Assumption \ref{ass:latent_rank} holds with $M f_{Y(1)|U}(\cdot |u) = \mu^1(u)$, and $U_i$ admits a new interpretation: $U_i$ is the rank of unit $i$ in terms of the systemic part of their treated potential outcome $\mu^1(U_i)$. Since $U_i$ also pins down the rank in terms of the systemic part of their untreated potential outcome $\mu^0(U_i)$ as well, the latent variable $U_i$ can be thought of as a `latent rank,' a single latent variable that determines the ranks of the systemic parts of both potential outcomes. Importantly, this does not imply that the two potential outcomes $Y_i(1)$ and $Y_i(0)$ be deterministically connected; only $\mu^1(U_i)$ and $\mu^0(U_i)$ are connected through $U_i$. In this sense, I refer to the conditional independence framework of this paper as `latent rank invariance,' which can be viewed as a direct relaxation of the rank invariance condition from the quantile treatment effect/IV literature: \citet{CH2005,CH2006,AI2006,VX2017,CL2019,HX2023}.

	Theorem \ref{thm:id} formally states the identification result. 

	\begin{theorem}\label{thm:id}
		Either Assumptions \ref{ass:assignment}-\ref{ass:id} or Assumptions \ref{ass:assignment}-\ref{ass:condind}, \ref{ass:id_cont}-\ref{ass:latent_rank} hold. Then, the joint density of $\big( Y_i(1), Y_i(0), D_i, X_i, Z_i \big)$ is identified. 
	\end{theorem}
	
	\begin{proof}
		See Appendix \ref{apdx:thm:id}.
	\end{proof}
	
	\noindent Another interpretation of Theorem \ref{thm:id} under Assumption \ref{ass:id} is that it is a point identification adaptation of the set identification result from \citet{HKS2014}; the additional identifying power comes from the conditional independence between $Y_i(d)$ and $X_i$ given $U_i$. 
		
	\section{Implementation}\label{sec:est}
	
	In implementation, I develop an estimation strategy under the finite support assumption on $U_i$.\footnote{Alternatively, we can construct a sieve maximum likelihood estimator, as suggested in the nonclassical measurement error literature, and let $U_i$ be continuous. The specifics of the sieve MLE are discussed in the Supplemental Appendix.} The focus on the case of discrete $U_i$ has several reasons. Firstly, discretization is often used in econometric models with latent heterogeneity as an approximation to a continuous latent heterogeneity space: see \citet{bonhomme2022discretizing} for more. Secondly, with parametrization, the estimation of infinite-dimensional objects such as conditional densities $f_{U|D=0,Z}$ and $f_{U|D=1,Z}$ becomes estimation of finite-dimensional objects $\Lambda_0$ and $\Lambda_1$. Lastly, with discrete $U_i$, the DTE parameters, which are nonlinear functionals of the densities $\big( f_{Y(1)|U}, f_{Y(0)|U}, f_U \big)$, becomes linear in quantities identified with quadratic moments: Proposition \ref{prop:gmm}.
	
	The last merit is particularly pertinent since it helps develop an easy-to-implement inferential procedure for the DTE parameters. Under the finite support assumption, an asymptotic linear representation of a DTE estimator  becomes a weighted sum of $U$-statistics of order 2. Since only the weights depend on first-step nuisance parameter estimators, inference on DTE parameters becomes straightforward, through the $U$-statistics asymptotic theory and Neyman orthogonalization. This procedure circumvents computationally heavy bootstrapping and/or an analytically complicated delta method, and is widely applicable to moment-based estimators in a finite mixture model. 

	All of the discussions in this section assume that $K$, the number of points in the support of $U_i$, is known. In practice, we often do not have \textit{a priori} choice of $K$. Thus, in Appendix \ref{appsec:choiceK}, I discuss how to apply the existing econometric methods such as the eigenvalue ratio estimator from \citet{AH2013} and the rank test from \citet{KP2006} for guidance on the choice of $K$. 

	Since I focus on discrete $U_i$, all of the statistical results in this section are developed under Assumptions \ref{ass:assignment}-\ref{ass:id}. In addition to the partitions $\{\mathcal{X}^m\}_{m=1}^{M_X}$ and $\{\mathcal{Z}^j\}_{j=1}^{M_Z}$ from Assumption \ref{ass:id}.b, I assume that there exists a partition $\{\mathcal{Y}^m\}_{m=1}^{M_Y}$ on $\mathbb{R}$ such that for any $u \neq u'$, Assumption \ref{ass:id}.c holds with some $\mathcal{Y} \in \{\mathcal{Y}^m\}_{m=1}^{M_Y}$. Also, I collapse the partition on $Z_i$ so that $M_Z = K$; Assumption \ref{ass:id}.b implies that both $\Lambda_0$ and $\Lambda_1$ are invertible.

	\subsection{Feasible moment for distributional treatment effect}\label{subsec:feasible_moment}

	Suppose that a DTE parameter of interest is identified by a moment function $m$ defined on $\big( Y_i(1), Y_i(0), D_i, X_i \big)$:\footnote{The fact that the moment condition \eqref{eq:original_moment} does not use $Z_i$ is compatible with the general identification result of Theorem \ref{thm:id}. We can extend Proposition \ref{prop:gmm} so that $m$ in \eqref{eq:original_moment} also takes $Z_i$ as an input, by using higher-order $U$-statistics.} 
	\begin{align}
	\E \left[ m \big( Y_i(1), Y_i(0), D_i, X_i ; \theta \big)\right] = 0. \label{eq:original_moment}
	\end{align}
	Each of the DTE parameters discussed in Section \ref{subsec:DTE} is applicable here. From Theorem \ref{thm:id}, $\theta$ is identified under Assumptions \ref{ass:assignment}-\ref{ass:id}. However, we cannot directly use \eqref{eq:original_moment} to estimate $\theta$, since we do not observe $Y_i(1)$ and $Y_i(0)$ simultaneously. Thus, we need to modify this infeasible moment condition into a feasible one.  

	To find a feasible moment condition, I construct a factorization similar to \eqref{eq:finite_mixture}, using conditional density of $Y_i$ instead of conditional probability of $(Y_i, X_i)$. Stacking the factorization across $\{\mathcal{Z}^j\}_{j=1}^K$ for some fixed $d \in \{0,1\}$, we get \begin{align*} 
	\Big( f_{Y|D=d,Z}(y|\mathcal{Z}^1) \, \cdots \, f_{Y|d,Z}(y|\mathcal{Z}^{K}) \Big) = \Big( f_{Y(d)|U}(y|u^1) \, \cdots \, f_{Y(d)|U}(y|u^K) \Big) \Lambda_d.
	\end{align*}
	Under Assumption \ref{ass:id}.b, $\Lambda_d$ is invertible: \begin{align*} 
	\Big( f_{Y(d)|U}(y|u^1)  \cdots  f_{Y(d)|U}(y|u^K) \Big) = \Big( f_{Y|D=d,Z}(y|\mathcal{Z}^1) \cdots f_{Y|D=d,Z}(y|\mathcal{Z}^{K}) \Big) \big( \Lambda_d \big)^{-1}.
	\end{align*}
	Let $\tilde{\Lambda}_d = \left( \Lambda_d \right)^{-1}$ for $d=0,1$ and let \begin{align*}
	p_{D,U} (d,k) &= \Pr \left\lbrace D_i=d, U_i = u^k \right\rbrace \hspace{5mm} \text{for } d=0,1 \text{ and } k=1, \cdots, K, \\
	p_{D,Z}(d,j) &= \Pr \left\lbrace D_i=d, Z_i \in \mathcal{Z}^j \right\rbrace \hspace{5mm} \text{for } d=0,1 \text{ and } j=1, \cdots, K.
	\end{align*}
	Then, with $\tilde{\lambda}_d(j,k)$ denoting the $j$-th row, $k$-th column entry of $\tilde{\Lambda}_d$ and $f_{Y,D=d,Z}$ denoting the joint density of $(Y_i,D_i,Z_i)$, \begin{align}
	f_{Y(d)|U}(y|u^k) = \sum_{j=1}^K \tilde{\lambda}_d(j,k) \cdot f_{Y|D,Z}(y|d,\mathcal{Z}^j) = \sum_{j=1}^K \frac{\tilde{\lambda}_d(j,k)}{p_{D,Z}(d,j)} \cdot f_{Y,D=d,Z}(y,\mathcal{Z}^j). \label{eq:linear_substitution}
	\end{align}
	The conditional density of latent variables is a linear combination of observed densities where the weight depend on $\tilde{\lambda}_d$ and $p_{D,Z}$. 

	Proposition \ref{prop:gmm} uses \eqref{eq:linear_substitution} and identifies the parameter of interest $\theta$ with a feasible quadratic moment condition. 

	\begin{proposition}\label{prop:gmm} Assumptions \ref{ass:assignment}-\ref{ass:id} hold. Suppose that a parameter of interest $\theta$ is identified by an infeasible moment condition \eqref{eq:original_moment}. Then, there exists a symmetric moment function $\tilde{m}$ such that $\theta$ is identified by a feasible quadratic moment condition \begin{align}
	\E \left[ \tilde{m} \big( W_i, W_{i'}; \theta, \tilde{\lambda}, p \big) \right] = 0 \label{eq:modified_moment}
	\end{align}
	where $W_i = (Y_i, D_i, X_i, Z_i)$,  $W_i \independent W_{i'}$, $\tilde{\lambda}$ denotes the collection of $\tilde{\lambda}_d(j,k)$, and $p$ denotes the collection of $p_{D,U}(d,k)$ and $p_{D,Z}(d,j)$.
	\end{proposition}

	\begin{proof}
	See Appendix \ref{apdx:prop:gmm}. 
	\end{proof}

	\noindent For example, the marginal distribution of treatment effect $\theta = F_{Y(1)-Y(0)}(\delta)$ for a given $\delta$ is identified with \begin{align}
	\tilde{m} \big(W_i, W_{i'} \big) &= \sum_{j=1}^K \sum_{j'=1}^K \frac{\sum_{k=1}^K \big( p_{D,U}(0,k) + p_{D,U}(1,k) \big) \tilde{\lambda}_0(j,k) \tilde{\lambda}_1(j,k)}{p_{D,Z}(0,j) p_{D,z}(1,j')} \notag \\
	& \quad \cdot \left( \frac{\mathbf{1}\{Y_{i'} \leq Y_i + \delta, D_i=0, Z_i \in \mathcal{Z}^j, D_{i'}=1, Z_{i'} \in \mathcal{Z}^{j'} \}}{2} \right. \notag \\
	& \quad \left. \quad + \frac{\mathbf{1}\{Y_i \leq Y_{i'} + \delta, D_{i'}=0, Z_{i'} \in \mathcal{Z}^j, D_i = 1, Z_i \in \mathcal{Z}^{j'} \}}{2} \right) - \theta. \label{eq:example_modified_moment}
	\end{align}
	In Sections \ref{sec:sim}-\ref{sec:emp}, the marginal distribution of treatment effect is used as a working example of the DTE parameter. The proof of Proposition \ref{prop:gmm} is constructive; application of Proposition \ref{prop:gmm} to the other moment-identified DTE parameters is straightforward.

	\subsection{Nonparametric finite mixture estimation}\label{subsec:NMF}

	To use the feasible moment condition \eqref{eq:modified_moment}, we need to first estimate the nuisance parameter $\big( \tilde{\lambda}, p \big)$. Find that $\big(\tilde{\lambda}, p \big)$ are functions of the mixture weights in the mixture representation \eqref{eq:finite_mixture}; $\tilde{\lambda}$ comes from the inverse of $\Lambda_d$, the mixture weights matrix, and $p_{D,U} (d,k) = \sum_{j=1}^K \lambda_d(k|j) \cdot p_{D,Z}(d,j)$. Thus, it suffices to estimate the (finite) mixture model to estimate the nuisance parameters. 
	
	To this end, I propose a new estimator for the nonparametric finite mixture model, which minimizes a quadratic loss between observed conditional probabilities and their model predictions: a nonlinear least-square estimator. To construct the least-squares objective, I take the given partition $\{\mathcal{Y}\}_{m=1}^{M_Y}$, $\{\mathcal{X}\}_{m=1}^{M_X}$ and $\{\mathcal{Z}\}_{j=1}^K$ and estimate the (discretized) conditional probability \[
	h(m,m'|d,j) := \Pr \left\lbrace Y_i \in \mathcal{Y}^m, X_i \in \mathcal{X}^{m'} \, \big| \, D_i=d, Z_i \in \mathcal{Z}^j \right\rbrace
	\]
	with $
	\hat{h}(m,m'|d,j):= \frac{\sum_{i=1}^n \mathbf{1}\left\lbrace Y_i \in \mathcal{Y}^m, X_i \in \mathcal{X}^{m'}, D_i=d, Z_i \in \mathcal{Z}^j \right\rbrace}{\sum_{i=1}^n \mathbf{1}\left\lbrace D_i=d, Z_i \in \mathcal{Z}^j \right\rbrace}
	$
	for every $m \in \{1,\dots,M_Y\}, m' \in \{1,\dots, M_X\}, d \in \{0,1\}$ and $j \in \{1,\dots,K\}$. Let \begin{align*}
	\gamma_{Y(d)}(m|k) &:= \Pr \left\lbrace Y_i(d) \in \mathcal{Y}^m \, | \, U_i = u^k \right\rbrace \quad \forall d=0,1 \\
	\gamma_{X}(m'|k) &:= \Pr \left\lbrace X_i \in \mathcal{X}^{m'} \, | \, U_i = u^k \right\rbrace \\
	\lambda_{d}(k|j) &:= \Pr \left\lbrace U_i = u^k \, | \, D_i=d, Z_i \in \mathcal{Z}^j \right\rbrace \quad \forall d=0,1
	\end{align*}
	denote the model parameters of the finite mixture model; $\gamma_{Y(0)}, \gamma_{Y(1)}, \gamma_X$ are the mixture component probabilities and $\lambda_0, \lambda_1$ are the mixture weights. Then, the nonlinear least-square finite mixture estimator $\left( \hat{\gamma}_{Y(0)}, \hat{\gamma}_{Y(1)}, \hat{\gamma}_X, \hat{\lambda}_0, \hat{\lambda}_1 \right)$ solves \begin{align}
	\min_{\gamma,\lambda} \sum_{d \in \{0,1\}} \sum_{m=1}^{M_Y} \sum_{m'=1}^{M_X} \sum_{j=1}^K \left( \hat{h}(m,m'|d,j) - \sum_{k=1}^K \gamma_{Y(d)}(m|k) \cdot \gamma_X(m'|k) \cdot \lambda_d (k|j) \right)^2 \label{eq:NMF}
	\end{align}
	subject to the nonnegativity constraints and the sum-to-one constraints \begin{align*}
	\gamma_{Y(0)}(m|k), \gamma_{Y(1)}(m|k) &\geq 0  \quad \forall m =1,\dots, M_Y \text{ and } k=1,\dots,K, \\
	\gamma_{Y(1)}(m|k) &\geq 0 \quad \forall m =1,\dots, M_X \text{ and } k=1,\dots,K, \\
	\lambda_0(k|j), \lambda_1(k|j) &\geq 0 \quad \forall j =1,\dots, K \text{ and } k=1,\dots,K, \\
	\sum_{m=1}^{M_Y} \gamma_{Y(0)}(m|k) = \sum_{m=1}^{M_Y} \gamma_{Y(1)}(m|k) &= \sum_{m=1}^{M_X} \gamma_X(m|k) = 1 \quad \forall k=1,\dots, K, \\
	\sum_{k=1}^K \lambda_0(k|j) &= \sum_{k=1}^K \lambda_1(k|j) = 1 \quad \forall j=1,\dots,K.
	\end{align*}
	The multiplicative structure in \eqref{eq:NMF} makes the objective function nonlinear. 

	The objective function in \eqref{eq:NMF} uses a quadratic loss, rather than a likelihood-based loss such as Kullback-Leibler divergence. A crucial merit of the quadratic loss is that once we fix two out of the three sets of parameters $(\gamma_{Y(0)},\gamma_{Y(1)})$, $\gamma_X$, $(\lambda_0, \lambda_1)$, the objective becomes quadratic in the remainder. Since the nonnegativity and sum-to-one constraints are all linear, this allows us to solve \eqref{eq:NMF} using a block coordinate descent algorithm rotating through $(\gamma_{Y(0)},\gamma_{Y(1)}), \gamma_X$ and $(\lambda_0, \lambda_1)$, where each update is a quadratic program readily solvable with standard quadratic program solvers in most statistical software.\footnote{Any quadratic program solver that admits a positive-semidefinite Hessian matrix can be used.}  To find the global minimum, I consider various initial values $\left( \gamma_{Y(0)}, \gamma_{(1)}, \gamma_X \right)$.\footnote{To initialize $\left( \gamma_{Y(0)}, \gamma_{(1)}, \gamma_X \right)$, I use weighted averages of $\{\hat{h}(m,m'|d,j)\}_{m,m'}$ across $(d,j)$, with randomly drawn weights that sum to one.}

	Nonparametric estimation of a finite mixture model is often implemented through diagonalization: \citet{H2008,BJR}, among others. While the estimators based on diagonalization impose the sum-to-one constraints on the mixture weights $\lambda_0, \lambda_1$ and the mixture component probabilities $\gamma_{Y(0)}, \gamma_{Y(1)}, \gamma_X$ with normalization, they do not impose the nonnegativity constraints on them. In contrast, the nonlinear least-squares estimator \eqref{eq:NMF} explicitly impose both sum-to-one constraints and nonnegativity constraints, fully exploiting the available information from the finite mixture model. In Section \ref{sec:sim}, simulation studies show that the nonlinear least-square estimator outperforms a joint diagonalization-based estimator in finite sample. 

	Theorem \ref{thm:est} establishes the $\sqrt{n}$-consistency of the finite mixture estimators.
	\begin{theorem}\label{thm:est}
		Assumptions \ref{ass:assignment}-\ref{ass:id} hold. With probability going to one, there exists some permutation $\sigma$ on $\{1,\dots,K\}$ such that as $n \to \infty$, \begin{align*}
			\hat{\gamma}_{Y(d)}(m|k) &= \gamma_{Y(d)}(m|\sigma(k)) + O_p \left( n^{-\frac{1}{2}} \right), \\
			\hat{\gamma}_{X}(m'|k) &= \gamma_{X}(m'|\sigma(k)) + O_p \left( n^{-\frac{1}{2}} \right), \\
			\hat{\lambda}_d(k|j) &= \lambda_d(\sigma(k)|j) + O_p \left( n^{-\frac{1}{2}} \right),
		\end{align*}
		for $d \in \{0,1\}$, $m = 1, \dots, M_Y$, $m'=1, \dots, M_X$, $j=1,\dots,K$ and $k=1,\dots,K$. 
	\end{theorem}
	
	\begin{proof}
		See Appendix \ref{apdx:thm:est}.
	\end{proof}

	\noindent The nuisance parameter $\tilde{\lambda}$ and $p$ are estimated by plugging in $\hat{\lambda}_0$ and $\hat{\lambda}_1$ for $\lambda_0$ and $\lambda_1$. The plug-in estimators $\hat{\tilde{\lambda}}$ and $\hat{p}$ are $\sqrt{n}$-consistent as well. 

	\subsection{Distributional treatment effect estimator}\label{subsec:TE}

	Given a feasible moment condition from Section \ref{subsec:feasible_moment} and the nuisance parameter estimators from Section \ref{subsec:NMF}, I now develop a plug-in estimator for a DTE parameter $\theta$. The key contribution of this section is that I propose a generalized orthogonalization procedure for a moment-based estimator in a nonparametric finite mixture model. 
	
	Developing an asymptotic theory for the naive plug-in estimator via the delta method is analytically complicated when the constraints are binding and the nuisance parameter estimator may not admit a standard asymptotic linear representation. This is particularly relevant since the nonlinear least-square estimator \eqref{eq:NMF} explicitly imposes nonnegativity constraints, which often bind in sample. Thus, as an alternative remedy for the first-step estimation error, I orthogonalize the moment function. 
	
	Even though the nuisance parameter estimators $\hat{\tilde{\lambda}}$ and $\hat{p}$ are complex nonlinear functions of the estimated probabilities $\hat{h}(\cdot | \cdot)$, we know that $(\tilde{\lambda},p)$ satisfies the following equations at their true values: for any $(\mathcal{Y},d,\mathcal{X},k)$,\begin{align}
	\sum_{j=1}^{K} \frac{\tilde{\lambda}_d(j,k)}{p_{D,Z}(d,j)} &\Pr \left\lbrace Y_i \in \mathcal{Y}, X_i \in \mathcal{X}, D_i=d, Z_i \in \mathcal{Z}^j \right\rbrace \notag \\
	& = \sum_{j=1}^{K} \frac{\tilde{\lambda}_d(j,k)}{p_{D,Z}(d,j)} \Pr \left\lbrace Y_i \in \mathcal{Y}, D_i=d, Z_i \in \mathcal{Z}^j \right\rbrace \notag \\
	& \qquad \cdot \sum_{j'=1}^{K} \frac{\tilde{\lambda}_d(j',k)}{p_{D,Z}(d,j')} \Pr \left\lbrace X_i \in \mathcal{X}, D_i=d, Z_i \in \mathcal{Z}^{j'} \right\rbrace \label{eq:phiA} \\
	\Pr \left\lbrace D_i=d, X_i \in \mathcal{X} \right\rbrace &= \sum_{k=1}^K p_{D,U}(d,k) \sum_{j=1}^{K} \frac{\tilde{\lambda}_d(j,k)}{p_{D,Z}(d,j)} \Pr \left\lbrace X_i = x, D_i=d, Z_i \in \mathcal{Z}^j \right\rbrace. \label{eq:phiB}
	\end{align}
	Equation \eqref{eq:phiA} comes from Assumption \ref{ass:condind} that $Y_i(d) \independent X_i \text{ } | \text{ } U_i$
	and Equation \eqref{eq:phiB} comes from the law of iterated expectation.\footnote{$\Pr \{Y_i(d) \in \mathcal{Y}, X_i \in \mathcal{X} | U_i=u\} = \Pr \{ Y_i(d) \in \mathcal{Y} | U_i = u\} \cdot \Pr \{X_i \in \mathcal{X} | U_i=u\}$ and $\Pr \{D_i=d, X_i=x\} = \sum_{k=1}^K p_{D,U}(d,k) \Pr \{X_i=x|U_i=u^k\}$.} Note that Equation \eqref{eq:phiA} can be written as a quadratic moment condition and Equation \eqref{eq:phiB} as a linear moment condition. 
	
	By evaluating \eqref{eq:phiA} across $\mathcal{Y} \in \{\mathcal{Y}^m\}_{m=1}^{M_Y}$, $d \in \{0,1\}$, $\mathcal{X} \in \{\mathcal{X}^{m'}\}_{m'=1}^{M_X}$ and $k \in \{1,\dots,K\}$, we have $2 M_Y M_X K$ moments. Similarly, by evaluating \eqref{eq:phiB} across $d \in \{0,1\}$ and $\mathcal{X} \in \{\mathcal{X}^{m'}\}_{m'=1}^{M_X}$, we have $2 M_X$ moments. Using trivial moments for $p_{D,Z}$ that \begin{align}
	p_{D,Z}(d,j) = \Pr \left\lbrace D_i=d, Z_i \in \mathcal{Z}^j \right\rbrace, \label{eq:phiC}
	\end{align}
	we have $2 M_Y M_X K + 2 M_X + 2 K$ auxiliary moments in total. Let $\phi(W_i, W_{i'}; \tilde{\lambda}, p)$ denote the collection of these auxiliary moment functions after symmetrization. Then, $\E[\phi(W_i, W_{i'}; \tilde{\lambda}, p)] = 0$. To use $\phi$ in the orthogonalization, I show that the Jacobian matrix of $\phi$ has full rank.

	\begin{lemma}\label{lemma:jacobian}
	Assumptions \ref{ass:assignment}-\ref{ass:id} hold. Then, the following Jacobian matrix is full row rank: \begin{align}
	\begin{pmatrix} \E \left[ \frac{\partial}{\partial \tilde{\lambda}} \phi (W_i,W_{i'}; \tilde{\lambda}, p)\right] \\
	\E \left[ \frac{\partial}{\partial p} \phi (W_i,W_{i'}; \tilde{\lambda}, p)\right] \end{pmatrix} \in \mathbb{R}^{(2K^2 + 2K + 2K) \times (2 M_Y M_X K + 2M_X + 2K)}. \label{eq:jacobi}
	\end{align}
	\end{lemma}
	\noindent The proof is provided in the Supplemental Appendix.\footnote{An implication from the proof of Lemma \ref{lemma:jacobian} is that the Jacobian matrix in Lemma \ref{lemma:jacobian} is not full rank when $\Lambda_0, \Lambda_1$ have more than $K$ columns; there are too many nuisance parameters. Thus, the support of the discretized $Z_i$ has to be exactly $K$. Recall that Assumption \ref{ass:id}.b imposes that we have at least $K$ points in the support of $X_i$ and $Z_i$. Now, Lemma \ref{lemma:jacobian} imposes that whenever we have more than $K$ points in the support of $Z_i$, we need to collapse some part of the support to reduce it to $K$.} Then, with an additional nuisance parameter \begin{align*}
	\mu &= \begin{pmatrix} \E \left[ \frac{\partial}{\partial \tilde{\lambda}} \phi (W_i,W_{i'}; \tilde{\lambda}, p)\right] \\
	\E \left[ \frac{\partial}{\partial p} \phi (W_i,W_{i'}; \tilde{\lambda}, p)\right] \end{pmatrix}^+ 
	\begin{pmatrix} \E \left[ \frac{\partial}{\partial \tilde{\lambda}} \tilde{m} (W_i,W_{i'}; \tilde{\lambda}, p)\right] \\
	\E \left[ \frac{\partial}{\partial p} \tilde{m} (W_i,W_{i'}; \tilde{\lambda}, p)\right] \end{pmatrix} 
	\end{align*}
	\noindent with $A^+$ denoting the right-inverse of a full-row-rank matrix $A$, the moment function \begin{align*}
	\psi (W_i,W_{i'}; \theta, \tilde{\lambda}, p, \mu) &= \tilde{m} (W_i,W_{i'}; \theta, \tilde{\lambda}, p) - {\mu}^\intercal \phi(W_i,W_{i'}; \tilde{\lambda}, p)
	\end{align*}
	satisfies the first-order Neyman orthogonality. 
	
	Given some weighting matrix $A$ and the plug-in, sample analogue estimator $\hat{\mu}$ for $\mu$, the DTE estimator $\hat{\theta}$ is \[
	\hat{\theta} = \arg \min_{\theta} \bar{\psi}_n (\theta)^\intercal A \bar{\psi}_n (\theta)
	\]
	with $
	\bar{\psi}_n ( \theta ) = \binom{n}{2}^{-1} \sum_{i<i'} \psi \left( W_i, W_{i'}; \theta, \hat{\tilde{\lambda}}, \hat{p}, \hat{\mu} \right).
	$
	Theorem \ref{thm:clt} establishes the asymptotic normality of the DTE estimator. 

	\begin{theorem}\label{thm:clt}
		Assumptions \ref{ass:assignment}-\ref{ass:id}, \ref{ass:regular} hold. Then, \begin{align*}
		\sqrt{n} \left( \hat{\theta} - \theta \right) &\xrightarrow{d} \mathcal{N} \left( 0, \Sigma_\theta \right) 
		\end{align*}
		as $n \to \infty$ with some consistently estimable $\Sigma_\theta$.
	\end{theorem}

	\begin{proof}
		See Appendix \ref{apdx:thm:clt}
	\end{proof}

	\noindent Assumption \ref{ass:regular} is the regularity condition on the infeasible moment condition \eqref{eq:original_moment} and provided in Appendix \ref{apdx:regular}. The expressions for the asymptotic variance $\Sigma_\theta$ and a consistent plug-in estimator $\widehat{\Sigma}_\theta$ are provided in Appendix \ref{apdx:thm:clt}. 

	The procedure in Sections \ref{subsec:feasible_moment} and \ref{subsec:TE} applies to any moment-identified parameter $\theta$ in a nonparametric finite mixture model, not only to the DTE parameters considered here. The only requirement is that $\theta$ be moment-identified with a moment function that takes observed variables and the latent class/type variable, i.e., $U_i$ in this paper's setup. Then, we can substitute the joint density of the observed variables and the latent class/type variable with a linear combination of observed densities using Proposition \ref{prop:gmm}, and orthogonalize the feasible moment using Lemma \ref{lemma:jacobian}.
	
	Consequently, the procedure extends immediately to other settings where a finite mixture serves as a device for identification: for example, random coefficient models where the latent type models unobserved heterogeneity in coefficients (\citet{bajari2007linear,fox2011simple} among others), and dynamic discrete choice models with unobserved state variables (\citet{AM2011,KS2009} among others). In each of these settings, if a researcher estimates the finite mixture weights using a $\sqrt{n}$-consistent first-step estimator, they can apply Proposition \ref{prop:gmm} and Lemma \ref{lemma:jacobian} to obtain $\sqrt{n}$-asymptotically normal inference for their parameter of interest. This is particularly useful when the first step is a constrained, non-smooth $M$-estimation problem where standard delta method arguments do not apply nor an asymptotic linear representation is readily available.

	\section{Simulation}\label{sec:sim}

	In this section, I present Monte Carlo simulation results. I generated 2,000 random samples $\{Y_i, D_i, X_i, Z_i \}_{i=1}^n$, from two data generating processes (DGP) that satisfy Assumption \ref{ass:assignment}-\ref{ass:id}. The proxy variables and the latent variable $X_i, Z_i, U_i$ are discrete with three points in their supports: $X_i, Z_i, U_i \in \{1,2,3\}$. In the first DGP, I let $(\Pr \{U_i=~1\}, \Pr \{U_i=2\}, \Pr \{U_i=3\}) = (0.3, 0.4, 0.3)$ and \begin{align*}
	\scalebox{0.98}{$\begin{pmatrix} \Pr \left\lbrace X_i =1 | U_i=k \right\rbrace \\ 
		\Pr \left\lbrace X_i =2 | U_i=k \right\rbrace \\
		\Pr \left\lbrace X_i =3 | U_i=k \right\rbrace
	\end{pmatrix} = \begin{pmatrix} \Pr \left\lbrace Z_i =1 | U_i=k \right\rbrace \\ 
		\Pr \left\lbrace Z_i =2 | U_i=k \right\rbrace \\
		\Pr \left\lbrace Z_i =3 | U_i=k \right\rbrace
	\end{pmatrix} = \begin{cases}
	\left( 0.756, 0.167, 0.078 \right)^\intercal & \text{if } k = 1 \\
	\left( 0.133, 0.733, 0.133 \right)^\intercal & \text{if } k = 2 \\
	\left( 0.078, 0.167, 0.756 \right)^\intercal & \text{if } k = 3
	\end{cases}.$}
	\end{align*}
	The potential outcomes $Y_i(1), Y_i(0)$ are normally distributed, conditioning on $U_i$. \[
	Y_i(d) \, \big| \, U_i=k \sim \mathcal{N} \big( \mu^k(d), {\sigma^k(d)}^2 \big)
	\]
	for $d=0,1$ where $\big( \mu^k(0), \sigma^k(0) \big)$ is (0,1), (1,1), (2,1) and $\big( \mu^k(1), \sigma^k(1) \big)$ is (1.5,1), (2,1), (2.5,0.5), respectively for $k=1,2,3$. Lastly, the treatment $D_i$ is drawn independently of $\big( Y_i(1), Y_i(0), X_i, Z_i, U_i \big)$, with $\Pr \{D_i=1\}=0.5$.

	In the second DGP, I only change the conditional distribution of $Z_i$ given $U_i$ to see how the estimation performance depends on the informativeness of the proxy $Z_i$: $$
	\begin{pmatrix} \Pr \left\lbrace Z_i =1 | U_i=k \right\rbrace \\ 
		\Pr \left\lbrace Z_i =2 | U_i=k \right\rbrace \\
		\Pr \left\lbrace Z_i =3 | U_i=k \right\rbrace
	\end{pmatrix} = \begin{cases}
	\left( 0.516, 0.323, 0.161 \right)^\intercal & \text{ if } k = 1 \\
	\left( 0.237, 0.526, 0.237 \right)^\intercal & \text{ if } k = 2 \\
	\left( 0.162, 0.323, 0.516 \right)^\intercal & \text{ if } k = 3
	\end{cases}.
	$$
	The total variation distance between the conditional distribution of $U_i$ given $Z_i$ and the marginal distribution of $U_i$, averaged over the marginal distribution of $Z_i$, i.e., \[
	TV := \sum_{z=1}^3 \frac{\Pr \{Z_i=z\}}{2} \cdot \sum_{u=1}^3 \left| \Pr \{U_i=u | Z_i=z\} - \Pr \{U_i=u\} \right|,
	\]
	is 0.407 in the first DGP and 0.182 in the second DGP. On average, observing $Z_i$ updates the conditional distribution of $U_i$ by a total variance distance of 0.407 in the first DGP and 0.182 in the second DGP.

	\begin{table}[t]
	\centering \small \renewcommand{\arraystretch}{1.3}
	\begin{tabular}{c|c|cccc|cccc}
    \hline
    $\delta$ & $F_{Y(1)-Y(0)}(\delta)$ & \multicolumn{4}{c|}{bias} & \multicolumn{4}{c}{rMSE} \\
	\hline \hline
	0 & 0.084 & -0.009 & 0.003 & -0.001 & 0.000 & 0.051 & 0.015 & 0.018 & 0.009 \\
	1 & 0.264 & -0.007 & 0.003 & 0.000 & 0.000 & 0.074 & 0.022 & 0.029 & 0.014 \\
	2 & 0.536 & 0.005 & 0.001 & -0.001 & -0.000 & 0.083 & 0.023 & 0.033 & 0.017 \\
	3 & 0.775 & 0.020 & 0.002 & 0.001 & -0.000 & 0.079 & 0.020 & 0.027 & 0.013 \\
	4 & 0.911 & 0.030 & 0.006 & 0.005 & 0.001 & 0.055 & 0.016 & 0.018 & 0.008 \\
	\hline
	\multicolumn{1}{c|}{$TV$} & & 0.182 & 0.182 & 0.407 & 0.407 & 0.182 & 0.182 & 0.407 & 0.407 \\
	\multicolumn{1}{c|}{$n$} & & 500 & 2000 & 500 & 2000 & 500 & 2000 & 500 & 2000 \\
	\hline
	\end{tabular}
	\caption{Bias and rMSE of DTE estimator $\widehat{F}_{Y(1)-Y(0)}(\delta)$.}
	\label{tab:sim_main} 
	\begin{minipage}{0.9\textwidth}
	\textit{Note:} The DTE estimators are trimmed to be between zero and one. 
	\end{minipage}
	\end{table}
	
	\begin{table}[t]
	\centering \small \renewcommand{\arraystretch}{1.3}
	\begin{tabular}{c|c|cccc}
    \hline
    $\delta$ & $F_{Y(1)-Y(0)}(\delta)$ & \multicolumn{4}{c}{coverage probability} \\
	\hline \hline
	0 & 0.084 & 0.990 & 0.992 & 0.966 & 0.961 \\
	1 & 0.264 & 0.994 & 0.992 & 0.960 & 0.961 \\
	2 & 0.536 & 0.992 & 0.989 & 0.964 & 0.951 \\
	3 & 0.775 & 0.992 & 0.990 & 0.959 & 0.956 \\
	4 & 0.911 & 0.987 & 0.980 & 0.946 & 0.947 \\
	\hline
	\multicolumn{1}{c|}{$TV$} & & 0.182 & 0.182 & 0.407 & 0.407 \\
	\multicolumn{1}{c|}{$n$} & & 500 & 2000 & 500 & 2000 \\
	\hline
	\end{tabular}
	\caption{Coverage of 95\% confidence interval.}
	\label{tab:sim_ci} 
	\begin{minipage}{0.9\textwidth}
	\textit{Note:} Standard error is computed with plug-in estimator for asymptotic variance.
	\end{minipage}
	\end{table}

	Given a simulated sample, I estimated the finite mixture using the nonlinear least-square estimator \eqref{eq:NMF}, with a three-way partition on $\mathbb{R}$ for $Y_i$,\footnote{I used $\{\mathcal{Y}^m\}_{m=1}^3 = \{(-\infty, 0.285], (0.285,1.9], (1.9,\infty)\}$. 0.285 and 1.9 are the 1/3-th and 2/3-th quantiles in the marginal distribution of $Y_i$.} and estimated the marginal distribution of treatment effect $F_{Y(1)-Y(0)}(\delta)$, evaluated at $\delta=0,1,\cdots,4$. Table \ref{tab:sim_main} contains the bias and the root mean squared error (rMSE) of the DTE estimators $\widehat{F}_{Y(1)-Y(0)}(\delta)$, across the two DGPs and sample size $n$. As $Z_i$ becomes more informative for $U_i$, the rMSE goes down. Table \ref{tab:sim_ci} shows that the coverage probability of the confidence intervals constructed with the asymptotic standard error achieves the target coverage and does so exactly when $Z_i$ is more informative. 
	
	\begin{table}[t]
	\centering \small \renewcommand{\arraystretch}{1.3}
	\begin{tabular}{c|c|cccc|cccc}
    \hline
    $\delta$ & $F_{Y(1)-Y(0)}(\delta)$ & \multicolumn{4}{c|}{bias} & \multicolumn{4}{c}{rMSE} \\
	\hline \hline
	0 & 0.084 & -0.000 & -0.003 & -0.007 & -0.002 & 0.054 & 0.023 & 0.025 & 0.010 \\
	1 & 0.264 & -0.004 & 0.000 & -0.003 & -0.001 & 0.076 & 0.024 & 0.031 & 0.014 \\
	2 & 0.536 & -0.008 & 0.007 & 0.001 & 0.000 & 0.085 & 0.027 & 0.035 & 0.016 \\
	3 & 0.775 & 0.004 & 0.009 & 0.007 & 0.002 & 0.083 & 0.026 & 0.031 & 0.014 \\
	4 & 0.911 & 0.018 & 0.006 & 0.012 & 0.003 & 0.055 & 0.019 & 0.024 & 0.010 \\
	\hline
	\multicolumn{1}{c|}{$TV$} & & 0.182 & 0.182 & 0.407 & 0.407 & 0.182 & 0.182 & 0.407 & 0.407 \\
	\multicolumn{1}{c|}{$n$} & & 500 & 2000 & 500 & 2000 & 500 & 2000 & 500 & 2000 \\
	\hline
	\end{tabular}
	\caption{Bias and rMSE of DTE estimator $\widehat{F}_{Y(1)-Y(0)}(\delta)$, with joint diagonalization.}
	\label{tab:sim_evd} 
	\begin{minipage}{0.9\textwidth}
	\textit{Note:} The DTE estimators are trimmed to be between zero and one. 
	\end{minipage}
	\end{table}

	\begin{table}[t]
	\centering \small \renewcommand{\arraystretch}{1.3}
	\begin{tabular}{c|cccc}
    \hline
    $(\tilde{\lambda}, p)$ estimation & \multicolumn{4}{c}{success rate} \\
	\hline \hline
	nonlinear LS & 0.918 & 0.999 & 1.000 & 1.000 \\
	joint diag. & 0.850 & 0.973 & 0.9795 & 1.000 \\
	\hline
	$TV$ & 0.182 & 0.182 & 0.407 & 0.407 \\
	$n$ & 500 & 2000 & 500 & 2000 \\
	\hline
	\end{tabular}
	\caption{Success rate for DTE estimation.}
	\label{tab:sim_time} \vspace{2mm}
	\begin{minipage}{0.9\textwidth}
	\textit{Note:} Success rate is proportion of simulated samples with completed estimation. 
	\end{minipage}
	\end{table}

	To compare, Table \ref{tab:sim_evd} contains the bias and the rMSE of the DTE estimator where the nuisance parameters $(\tilde{\lambda}, p)$ are estimated with a joint diagonalization estimator from \citet{BJR}. When $Z_i$ is less informative, the rMSE of the joint diagonalization-based estimator is up to 50\% larger. 
	
	Lastly, Table \ref{tab:sim_time} contains the estimation completion rates for the nonlinear least-square first-step and the joint diagonalization first-step. The nonlinear least-square estimator has noticeably higher completion rate when $Z_i$ is less informative and $n=500$; the joint diagonalization estimator produce numerically singular mixture weight estimates $\widehat{\Lambda}_0, \widehat{\Lambda}_1$ and Jacobian matrix \eqref{eq:jacobi} more often. Imposing the nonnegativity constraint improves finite sample performance of estimators based on a nonparametric finite mixture, both on the intensive margin and the extensive margin.

	\section{Empirical illustration}\label{sec:emp}
	
	As an empirical illustration, I revisit \citet{JMR} and estimate the effect of workplace wellness program on medical spending. As discussed in Example \ref{eg:panel} of Section \ref{sec:id}, \citet{JMR} fits the econometrics framework of this paper well. Firstly, the treatment, eligibility for a workplace wellness program, was randomly assigned. Secondly, the treatment mechanism is regime-changing in its nature since the workplace wellness program included information sessions on healthy lifestyle, which are designed to induce systemic changes in individuals' health-related behaviors including medical service-seeking and self-care practices. Lastly, the authors collected the outcome variable before the treatment assignment and after the treatment period, giving us two proxy variables.
	
	
	
	Taking advantage of the random assignment, \citet{JMR} estimated the intent-to-treat type ATE of the workplace wellness program on the monthly medical spending. The ATE estimate showed that the eligibility for the wellness program raised the monthly medical spending by \$10.8, with $p$-value of 0.937, finding no significant intent-to-treat effect. In \citet{JMR}, the authors acknowledge that the null effect on the mean does not necessarily mean null effect everywhere, though they themselves do not explore the treatment effect heterogeneity in the paper.\footnote{In the original dataset used in \citet{JMR}, the authors had connected the medical spending variables to additional survey variables such as age, health behavior, salary, etc. They did not explore how the treatment effect interacts with the additional characteristics, but they did add these additional control variables through double Lasso. Adding the control variables increased the point estimate for the ATE (\$34.9) but the estimate still remained insignificant, with $p$-value being 0.859.} On page 1890, \citet{JMR} state ``there may exist subpopulations who did benefit from the intervention or who would have benefited had they participated.''\footnote{Damon Jones, David Molitor, and Julian Reif, ``What do workplace wellness programs do? Evidence from the Illinois workplace wellness study,'' \textit{The Quarterly Journal of Economics}, vol. 134, no.4 (2019): 1747-1791.} I build onto this intuition and estimate the DTE parameters, to explore the treatment effect heterogeneity. 
	
	The dataset of \citet{JMR} contains monthly medical spending records for the following three time periods: July 2015-July 2016, August 2016-July 2017 and August 2017-January 2019. The experiment started in the summer of 2016 and the treated individuals were offered to participate in the wellness program starting the fall semester of 2016: \begin{align*}
	Y_i &: \text{ monthly medical spending during August 2016-July 2017} \\
	D_i &: \text{ a binary variable for whether eligible to participate in the wellness program} \\
	X_i &: \text{ monthly medical spending during July 2015-July 2016} \\
	Z_i &: \text{ monthly medical spending during August 2017-January 2019} 
	\end{align*}
	$X_i$ is the pretreatment outcome variable and $Z_i$ is the post-treatment outcome variable. To connect the hidden Markov model in Example \ref{eg:panel} to this empirical context, we can think of the common shock $V_{it}$ as underlying health status. The first-order Markovian assumption on the underlying health status is a common practice in health economics literature where the underlying health status variable is often modeled to be first-order autoregressive: \citet{grossman1972concept,wagstaff1993demand,jacobson2000family,yogo2016portfolio} and more. 

	In the finite mixture estimation step, I let $M_Y=4$ and $M_X=6$, using equal-probability bins based on quantiles. To choose $K$, the number of points in the support of $U_i$, I turn to the rank estimator/test from Appendix \ref{appsec:choiceK}. Both the rank estimator from \citet{AH2013} and the rank test from \citet{KP2006} suggest $K \geq 3$. Also, for all $K \geq 3$, the specification test from Appendix \ref{appsec:false} does not reject the null hypothesis that $X_i \independent D_i \; | \; U_i$. In the main text, I present the empirical results for $K=5$, the largest value that is not subject to overfitting concerns. Specifics on the choice of $K$ and the sensitivity analysis in terms of $K$ are provided in the Supplemental Appendix. 

	\begin{table}[t]
	\centering \small 
	\renewcommand{\arraystretch}{1.5}
	\begin{tabular}{cccccc}
	\hline
	& (1) & (2) & (3) & (4) & (5) \\
	\hline \hline
	$CATE(\mathcal{Y})$ (\$) & -20.42 & 164.26 & 297.43 & -157.34 & -732.82$^*$ \\
	& (166.28) & (388.97) & (350.05) & (515.44) & (442.98) \\
	\hline
	$\mathcal{Y}$ & $(-\infty, 42.7]$ & $(42.7, 132.2]$ & $(132.2, 286.8]$ & $(286.8, 671.1]$ & $(671.1, \infty]$ \\
	\hline
	\end{tabular}
	\caption{Conditional average treatment effect $\E[Y_i(1)-Y_i(0) | Y_i(0) \in \mathcal{Y}]$.}
	\label{tab:CATE} \vspace{2mm}
	\begin{minipage}{1\textwidth}
	\footnotesize \textit{Note:} The five conditioning sets on $Y_i(0)$ as chosen with $F_{Y(0)}^{-1}(1/5), \cdots, F_{Y(0)}^{-1}(4/5)$. \\
	The medical spending variable is measured in \$. A 10\% significance level is denoted with $*$.
	\end{minipage}
	\end{table}

	Table \ref{tab:CATE} contains the estimates for conditional ATE, using $Y_i(0)$ as the conditioning variable. The five conditioning sets are defined with five quintiles of $Y_i(0)$. Conditioning on the fifth quintile of $Y_i(0)$, the CATE estimate is statistically significant at 10\% significance level. This suggests that the treatment may reduce the monthly medical spending among individuals with high baseline medical spending, while having little impact on the average. 

	\begin{figure}[t]
	\centering
	\includegraphics[width=0.7\textwidth]{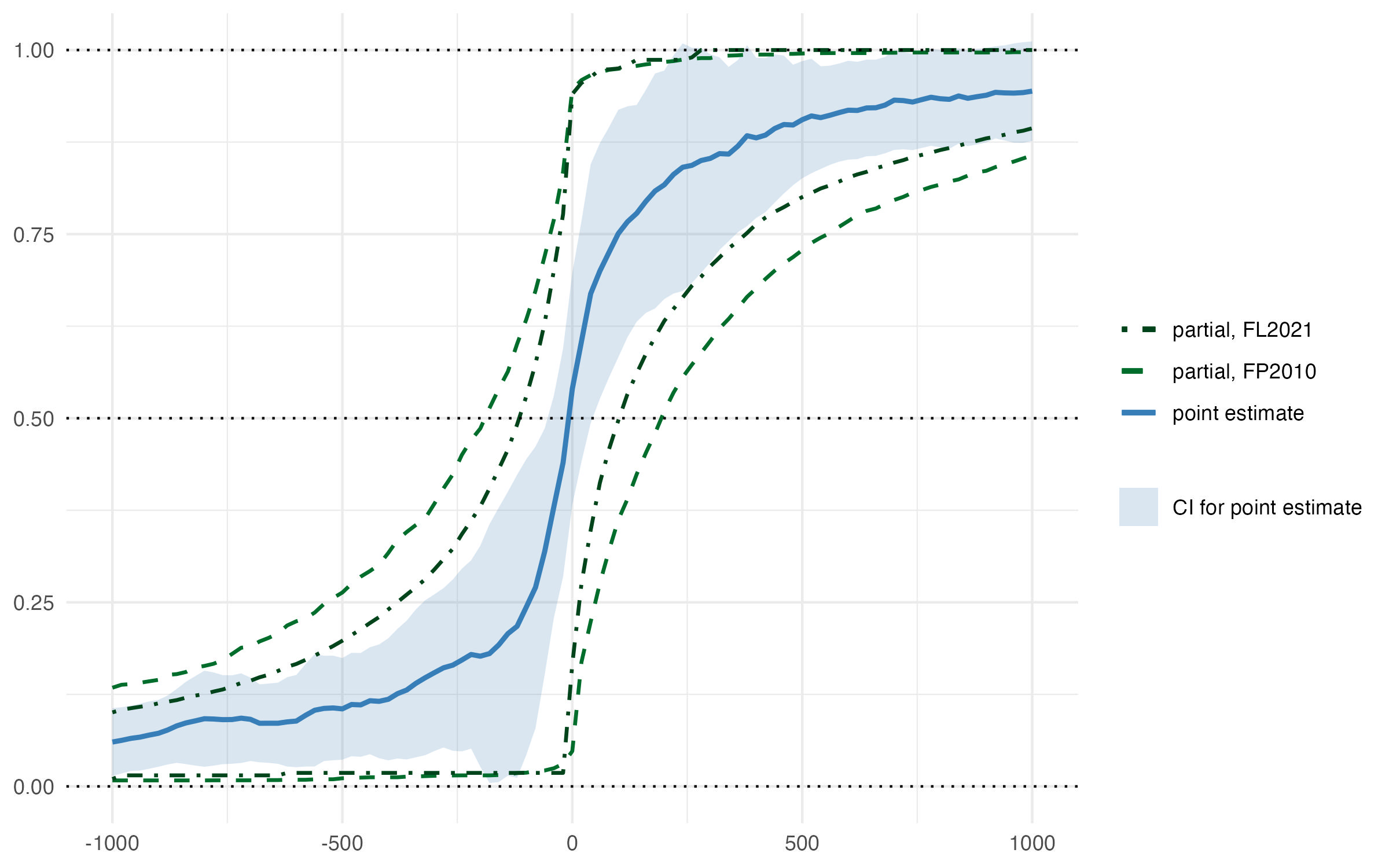}
	\caption{Marginal distribution of $Y_i(1) - Y_i(0)$.}
	\label{fig:marginal} \vspace{2mm}
	\begin{minipage}{1\textwidth}
	\footnotesize \textit{Note:} The distribution of treatment effect $Y_i(1)-Y_i(0)$ is estimated for $[-\$1000,\$1000]$. \\
	Additionally, 95\% pointwise confidence intervals and partial bounds from \citet{FP2010,FL2021} are provided. 
	\end{minipage}
	\end{figure}

	Figure \ref{fig:marginal} contains the estimated marginal distribution of the treatment effect and its 95\% pointwise confidence interval. Overall, it is unclear if the share of winners is bigger than 0.5. Though the point estimate $\widehat{F}_{Y(1)-Y(0)}(0)$ is 0.540, the 95\% confidence interval for $F_{Y(1)-Y(0)}(0)$ includes 0.5, not being able to reject the null $F_{Y(1)-Y(0)}(0)\leq 0.5$. As comparison, partial bounds from \citet{FP2010,FL2021} are plotted in Figure \ref{fig:marginal} as well. The DTE estimates are consistent with the partial identification, lying between the two bounds. At zero, the comparison highlights the information gain; the width of the 95\% confidence interval for $F_{Y(1)-Y(0)}(0)$ using the DTE estimator is 0.315 while the width of the two partial bounds is 0.774 and 0.952, respectively for \citet{FL2021} and \citet{FP2010}, providing little information on the share of winners.

	\section{Conclusion} 
	
	In this paper, I propose a conditional independence framework to identify and estimate the distributional treatment effect parameters. The point identification gives us much more information than the existing partial identification approaches. In developing the estimation strategy, this paper proposes a new estimator on a nonparametric finite mixture model which fully exploits the nonnegavitiy constraints on model parameters implied by the probability model.

	An important avenue for future research is how we extend the current estimation strategy to account for a continuous latent variable $U_i$, while retaining the desirable properties of the discretization-based estimation strategy, such as an easy inferential tool. For example, we may directly correct for the discretization bias by using some subsampling-based method such as the jackknife correction. Another important follow-up research direction would be to pursue a fully-developed policy learning framework based on the DTE estimation, providing distributional guarantees on policy recommendation.

	\clearpage 

	\appendix
	
	\begin{center}
		\LARGE{APPENDIX}
	\end{center}

	\section{Testable implication and specification test}\label{appsec:false}

	Suppose that $U_i$ is finitely discrete as in Assumption \ref{ass:id} and let $\gamma_{X,d}(m'|k)$ denote $\Pr \left\lbrace X_i \in \mathcal{X}^{m'} | D_i=d, U_i=u^k \right\rbrace$. Under Assumption \ref{ass:assignment}, we have testable implication
	\begin{align}
		\sum_{k=1}^K \sum_{m'=1}^{M_X} \left( \gamma_{X,0}(m'|k) - \gamma_{X,1}(m'|k) \right)^2 = 0. \label{eq:test}
	\end{align}
	To construct the test statistic based on \eqref{eq:test}, I modify the finite mixture estimator \eqref{eq:NMF}, by minimizing over two separate model parameters $\gamma_{X,0}$ and $\gamma_{X,1}$. Importantly, the two sets of model parameters $(\gamma_{Y(0)}, \gamma_{X,0}, \lambda_0)$ and $(\gamma_{Y(1)}, \gamma_{X,1}, \lambda_1)$ are still identified, but now up to two separate permutations on $\{u^1,\dots,u^K\}$. 
	
	Given the nuisance parameter estimator with treatment-status-specific $\gamma_{X,d}$, I construct an estimator for $\Pr \left\lbrace X_i \in \mathcal{X} | D_i=d, U_i=u \right\rbrace$, which applies the orthogonalization procedure from Section \ref{subsec:TE}: \begin{align*}
	\widehat{p}_{X|D=d,U}(\mathcal{X}|u^k) &= \binom{n}{2}^{-1} \sum_{i < i'} \left( \tilde{m} \big(W_i,W_{i'};\hat{\tilde{\lambda}},\hat{p} \big) + \mu^\intercal \phi \big(W_i, W_{i'};\hat{\tilde{\lambda}},\hat{p} \big) \right)
	\end{align*}
	where $\tilde{m} \big( W_i,W_{i'}; \tilde{\lambda}, p \big) = \sum_{j=1}^K \frac{\tilde{\lambda}_d(j,k)}{p_{D,Z}(d,j)} \cdot \frac{\mathbf{1}\{D_i=d, X_i \in \mathcal{X}, Z_i \in \mathcal{Z}^j\}  + \mathbf{1}\{D_{i'}=d, X_{i'} \in \mathcal{X}, Z_{i'} \in \mathcal{Z}^j\}}{2}$. For any arbitrary partition $\{\mathcal{X}^{m'}\}_{m'=1}^{\tilde{M}}$ on $\mathbb{R}$, Theorem \ref{thm:clt} extends to   \begin{align}
	T_n := & n \begin{pmatrix} \widehat{p}_{X|D=0,U}(\mathcal{X}^1|u^1) - \widehat{p}_{X|D=1,U}(\mathcal{X}^1|u^1) \\ \vdots \\ \widehat{p}_{X|D=0,U}(\mathcal{X}^{\tilde{M}}|u^K) - \widehat{p}_{X|D=1,U}(\mathcal{X}^{\tilde{M}}|u^K)\end{pmatrix}^\intercal {\widehat{\Sigma}_X}^{-1} \notag \\
	& \hspace{8mm} \cdot \begin{pmatrix} \widehat{p}_{X|D=0,U}(\mathcal{X}^1|u^1) - \widehat{p}_{X|D=1,U}(\mathcal{X}^1|u^1) \\ \vdots \\ \widehat{p}_{X|D=0,U}(\mathcal{X}^{\tilde{M}}|u^K) - \widehat{p}_{X|D=1,U}(\mathcal{X}^{\tilde{M}}|u^K)\end{pmatrix} \xrightarrow{d} \chi^2 \big(\tilde{M} \cdot K \big) \label{eq:test_stat1}
	\end{align}
	as $n \to \infty$, with a plug-in estimator $\widehat{\Sigma}_X$ for an asymptotic variance ${\Sigma}_X$. 
	
	In practice, the test statistic $T_n$ is infeasible since the permutation on $U_i$ from the treated subpopulation and that from the untreated subpopulation are not matched. Since $K$ is finite, I compute \eqref{eq:test_stat1} for every permutation on $\{1,\cdots,K\}$ and take the minimum. The test is still valid but conservative. In addition, although $\Sigma_X$ is nonsingular, certain rows of $\Sigma_X$ may be nearly collinear due to the linear constraint that $\sum_{m'=1}^{\tilde{M}} \Pr \{X_i \in \mathcal{X}^{m'} | D_i=d, U_i=u\}=1$ at truth, making the first-order asymptotic approximation to $\Sigma_X$ unreliable in finite sample. Thus, I drop $K$ entries in \eqref{eq:test_stat1} associated with $\mathcal{X}^{\tilde{M}}$ and use the limiting distribution $\chi^2 \big( (\tilde{M} -1) \cdot K \big)$. Table \ref{tab:sim_size} contains the simulation result on the feasible test, from the simulation exercise in Section \ref{sec:sim}. As expected, the test attains the intended significance level, and is conservative when the proxy variable $Z_i$ is less informative and $n$ is small. 

	\begin{table}[h]
    \centering \small \renewcommand{\arraystretch}{1.3}
	\begin{tabular}{c|cccc}
    \hline
    $\alpha$ & \multicolumn{4}{c}{size} \\
	\hline \hline
	0.05 & 0.038 & 0.018 & 0.026 & 0.042 \\
	0.10 & 0.061 & 0.035 & 0.056 & 0.092 \\
	\hline
	\multicolumn{1}{c|}{$TV_{U|Z, U}$} & 0.182 & 0.182 & 0.407 & 0.407 \\
	\multicolumn{1}{c|}{$n$} & 500 & 2000 & 500 & 2000 \\
	\hline
	\end{tabular}
    \caption{Size of the specfication test with significance level $\alpha$.}
    \label{tab:sim_size}
	\end{table}

	\section[Choice of K]{Choice of $K$ for a discrete $U_i$}\label{appsec:choiceK}

	To fix the number of components in the finite mixture, $K$, I use rank estimation and inference methods. Consider a $M_X \times 2 M_Z$ matrix $\mathbf{H}_X$: \begin{align*}
		\scalebox{0.97}{$\mathbf{H}_X = \begin{pmatrix} \Pr \left\lbrace X_i \in \mathcal{X}^1 | D_i = 0, Z_i \in \mathcal{Z}^1 \right\rbrace & \cdots & \Pr \left\lbrace X_i \in \mathcal{X}^1 | D_i = 1, Z_i \in \mathcal{Z}^{M_Z} \right\rbrace \\
		\vdots & \ddots & \vdots \\
		\Pr \left\lbrace X_i \in \mathcal{X}^{M_X} | D_i = 0, Z_i \in \mathcal{Z}^1 \right\rbrace & \cdots & \Pr \left\lbrace X_i \in \mathcal{X}^{M_X} | D_i = 1, Z_i \in \mathcal{Z}^{M_Z}) \right\rbrace \end{pmatrix}.$} 
	\end{align*}
	with some partitions $\{\mathcal{X}^{m'}\}_{m'=1}^{M_X}$ and $\{\mathcal{Z}^j\}_{j=1}^{M_Z}$. $\mathbf{H}_X$ pools information from the treated subpopulation and the untreated subpopulation. From Assumptions \ref{ass:assignment} and \ref{ass:id}, the rank of $\mathbf{H}_X$ is $K$. I estimate $\mathbf{H}_X$ with \begin{align*}
		\mathbb{H}_X &= \begin{pmatrix} \frac{\sum_{i=1}^n \mathbf{1}\left\lbrace D_i=0, X_i \in \mathcal{X}^1, Z_i \in \mathcal{Z}^1 \right\rbrace}{\sum_{i=1}^n \mathbf{1}\left\lbrace D_i=0, Z_i \in \mathcal{Z}^1 \right\rbrace} & \cdots & \frac{\sum_{i=1}^n \mathbf{1}\left\lbrace D_i=1, X_i \in \mathcal{X}^1, Z_i \in \mathcal{Z}^{M_Z}\right\rbrace}{\sum_{i=1}^n \mathbf{1}\left\lbrace D_i=1, Z_i \in \mathcal{Z}^{M_Z} \right\rbrace} \\
		\vdots & \ddots & \vdots \\
		\frac{\sum_{i=1}^n \mathbf{1}\left\lbrace D_i=0, X_i \in \mathcal{X}^{M_X}, Z_i \in \mathcal{Z}^1 \right\rbrace}{\sum_{i=1}^n \mathbf{1}\left\lbrace D_i=0, Z_i \in \mathcal{Z}^1 \right\rbrace} & \cdots & \frac{\sum_{i=1}^n \mathbf{1}\left\lbrace D_i=1, X_i \in \mathcal{X}^{M_X}, Z_i \in \mathcal{Z}^{M_Z}\right\rbrace}{\sum_{i=1}^n \mathbf{1}\left\lbrace D_i=1, Z_i \in \mathcal{Z}^{M_Z} \right\rbrace} \end{pmatrix}. 
	\end{align*}

	Firstly, to estimate $K$ using $\mathbb{H}_X$, I use the eigenvalue ratio estimator from \citet{AH2013}.\footnote{The eigenvalue ratio estimator is developed for a setup where a low-rank matrix of growing dimension is estimated, which is different from the setup of this paper; as $n$ increases, the dimension of $\mathbf{H}_X$ stays the same while the estimation error of $\mathbb{H}_X$ decreases. Following \citet{H2025}'s treatment, I apply the eigenvalue ratio estimator of \citet{AH2013} to $\mathbb{H}_X$.} 
	Let $\nu_k(\mathbf{H}_X)$ denote the $k$-th largest eigenvalue of $\mathbf{H}_X {\mathbf{H}_X}^\intercal$ when $M_X \leq 2 M_Z$ and that of ${\mathbf{H}_X}^\intercal \mathbf{H}_X$ when $M_X \geq 2 M_Z$. The eigenvalue ratio estimator is $$
	\widehat{K}_{ER} = \max_{K \leq K_{\max}} \frac{\nu_K(\mathbb{H}_X)}{\mu_{K+1}(\mathbb{H}_X)} 
	$$
	with $K_{\max} = \min \{M_X, 2M_Z\} -1$. Similarly, \citet{AH2013} also proposes an estimator based on the growth rate of eigenvalues: $$
	\widehat{K}_{GR} = \max_{K \leq K_{\max}} \frac{\log \left( 1 + \nu_K(\mathbb{H}_X) \big/ \sum_{k=K+1}^{M_X} \nu_k(\mathbb{H}_X) \right)}{\log \left( 1+ \nu_{K+1}(\mathbb{H}_X) \big/ \sum_{k=K+2}^{M_X} \nu_k(\mathbb{H}_X) \right)}
	$$
	with $K_{\max} = \min \{M_X, 2M_Z\}-2$. Both estimators are consistent. 

	Secondly, to infer on $K$, I use the Kleibergen-Paap (KP) rank test: \citet{KP2006}. For a given $K$, the KP rank test tests the null hypothesis $H_0: \text{rank} (\mathbf{H}_X) = K$ against the alternative hypothesis $H_1: \text{rank} (\mathbf{H}_X) \geq K+1$.\footnote{The KP rank test is developed for a general setup where a low-dimensional, low-rank matrix is estimated with estimators that are asymptotically normal. Since $\text{vec}(\mathbb{H}_X)$ is asymptotically normal around $\text{vec}(\mathbf{H}_X)$, we can directly apply the KP rank test.} The construction of the KP rank test statistic is implemented easily with singular value decomposition. 

	\section{Regularity conditions on the DTE moment}\label{apdx:regular}

	\begin{assumption}\label{ass:regular} Assume \begin{enumerate}[label=\textbf{\alph*.}]
	\item $\Theta$, the parameter space for $\theta$, is a compact subset of $\mathbb{R}^p$ with $p < \infty$. \\
	The true value of $\theta$, denoted with $\theta^0$, lies in the interior of $\Theta$.
	\item There exists some $\underline{\nu} > 0$ such that $\min \{ |\det(\Lambda_0)|, |\det(\Lambda_1)|\} \geq \underline{\nu}$. \\
	Thus, the parameter space for $\tilde{\lambda}$ is a compact subset of $\mathbb{R}^{2K^2}$.
	\item $\E \left[ m \big( Y_i(1), Y_i(0), D_i, X_i; \theta^0 \big) \right] = 0$ and for any $\varepsilon > 0$, \\
	$\inf_{\|\theta - \theta^0\|_2 \geq \varepsilon} \left\| \E \left[ m \big( Y_i(1), Y_i(0), D_i, X_i; \theta \big) \right] \right\|_2 > 0$.
	\item For every $d \in \{0,1\}$, $j=1, \dots, K$ and $j' = 1, \dots, K$, \\
	$\E \left[ \left| m (Y_{i'},Y_i,d,X_{i'},\theta^0) \mathbf{1}\{D_i=0,Z_i\in \mathcal{Z}^j, D_{i'}=1, Z_{i'} \in \mathcal{Z}^{j'}\} \right| \right]$ is finite and \begin{align*}
	\sup_{\theta \in \Theta} &\left\| \binom{n}{2}^{-1} \sum_{i < i'} m (Y_{i'},Y_i,d,X_{i'},\theta) \mathbf{1}\{D_i=0,Z_i\in \mathcal{Z}^j, D_{i'}=1, Z_{i'} \in \mathcal{Z}^{j'}\} \right. \\
	& \qquad \left. - \E \left[ m (Y_{i'},Y_i,d,X_{i'},\theta) \mathbf{1}\{D_i=0,Z_i\in \mathcal{Z}^j, D_{i'}=1, Z_{i'} \in \mathcal{Z}^{j'}\} \right] \right\|_2 \xrightarrow{p} 0.
	\end{align*}
	\item $\theta \mapsto m(Y_{i'}, Y_i, 0, X_{i'};\theta)$ and $\theta \mapsto m(Y_{i'}, Y_i, 1, X_{i'};\theta)$ are almost surely differentiable and for every $d \in \{0,1\}$, $j=1, \dots, K$ and $j' = 1, \dots, K$, \begin{align*}
	\sup_{\theta \in \Theta} &\left\| \binom{n}{2}^{-1} \sum_{i < i'} \frac{\partial}{\partial \theta} m (Y_{i'},Y_i,d,X_{i'},\theta) \mathbf{1}\{D_i=0,Z_i\in \mathcal{Z}^j, D_{i'}=1, Z_{i'} \in \mathcal{Z}^{j'}\} \right. \\
	& \; \; \left. - \E \left[ \frac{\partial}{\partial \theta} m (Y_{i'},Y_i,d,X_{i'},\theta) \mathbf{1}\{D_i=0,Z_i\in \mathcal{Z}^j, D_{i'}=1, Z_{i'} \in \mathcal{Z}^{j'}\} \right] \right\|_2 \xrightarrow{p} 0.
	\end{align*}
	\item The Jacobian matrix $
	\Psi := \E \left[ \frac{\partial}{\partial \theta} \psi \left( W_i, W_{i'}; \theta, \tilde{\lambda}^0, p^0, \mu^0 \right) \Big|_{\theta = \theta^0} \right]
	$
	is full rank. 
	\item Let $\tilde{\psi}(w;\theta,\tilde{\lambda},p,\mu) = \E[\psi(W_i,w;\theta,\tilde{\lambda},p,\mu)]$. \\
	$\E \left[ \tilde{\psi} \left( W_i;\theta^0, \tilde{\lambda}^0, p^0, \mu^0 \right) \tilde{\psi} \left( W_i;\theta^0, \tilde{\lambda}^0, p^0, \mu^0 \right)^\intercal \right]$ is finite and positive-definite. 
	\end{enumerate}
	\end{assumption}

	\section{Proofs}
		
	\subsection[Proof for Theorem 1]{Proof for Theorem \ref{thm:id}}\label{apdx:thm:id}
	
	The proof for Theorem \ref{thm:id} under Assumptions \ref{ass:assignment}-\ref{ass:id} is direct from Theorem 1 of \citet{H2008}. Applying the spectral decomposition of \citet{H2008} twice, firstly for the treated subpopulation and secondly for the untreated subpopulation, we have two diagonalization result. From Assumption \ref{ass:assignment}, the conditional distribution of $X_i$ given $U_i$ is shared across the two diagonalizations. Thus, we can connect the two permutations on $\{u^1, \dots, u^K\}$ using the columns of $\Gamma_X$, completing the proof. 
	
	To prove Theorem \ref{thm:id} under Assumptions \ref{ass:assignment}-\ref{ass:condind}, \ref{ass:id_cont}-\ref{ass:latent_rank}. we apply the spectral decomposition of \citet{HS2008} twice, which gives us the infinite collection $\left\lbrace f_{Y(1)|U}(\cdot|u), f_{Y(0)|U} (\cdot|u),  f_{X|U}(\cdot|u) \right\rbrace_{u \in \mathcal{U}}$, without a labeling on $u$. To find an ordering on the infinite number of triads, let $\tilde{U}_i = h(U_i) := M f_{Y(d)|U}( \cdot | U_i)$ from Assumption \ref{ass:latent_rank} and let $\tilde{\mathcal{U}} = h(\mathcal{U})$. Now, we have labeled each triad with $\tilde{u} = h(u)$ and therefore identified $f_{Y(1)|\tilde{U}}, f_{Y(0)|\tilde{U}}$ and $f_{X|\tilde{U}}$. Note that both Assumptions \ref{ass:assignment}-\ref{ass:condind} hold with $\tilde{U}_i$ since $h$ is strictly increasing. 
	
	To complete the proof, it suffices to show that the marginal distribution of $\tilde{U}_i$ is identified. For that, firstly I establish the injectivity of the integral operator based on the conditional density of $X_i$ given $\tilde{U}_i$. Find that $f_{X|\tilde{U}} (x|\tilde{u}) = f_{X|U} \left( x|h^{-1}(\tilde{u}) \right)$ and \begin{align*}
		\left[ L_{X|\tilde{U}} g \right] (x) &= \int_{\mathcal{\tilde{U}}} f_{X|\tilde{U}} (x|\tilde{u}) g(\tilde{u}) d \tilde{u} = \int_{\mathcal{\tilde{U}}} f_{X|U} \left( x|h^{-1}(\tilde{u}) \right) g(\tilde{u}) d \tilde{u} \\
		&= \int_{\mathcal{U}} f_{X|U} (x|u) g \left( h (u) \right) h'(u) d u, \hspace{5mm} \text{by letting } \tilde{u} = h(u).
	\end{align*}
	From the completeness of $f_{X|U}$, $L_{X|\tilde{U}} g = 0$ implies that $g(h(u)) h'(u) = 0$ for almost everywhere on $\mathcal{U}$. Since $h$ is strictly increasing, $h'(u) > 0$. Thus, we have $g(\tilde{u}) = 0$ almost everywhere on $\tilde{\mathcal{U}}$: the completeness of $f_{X|\tilde{U}}$ follows. Using the completeness, we identify $f_{\tilde{U}|D=d,Z}$ from $
	f_{X|D=d,Z} = \int_{\mathbb{R}} f_{X|\tilde{U}} (x|\tilde{u}) f_{\tilde{U}|D=d,Z} (\tilde{u}|z) d \tilde{u}. 
	$
	Since the conditional density of $X_i$ given $D_i, Z_i$ is observed, the marginal density of $\tilde{U}_i$ is identified. \qed 

	\subsection[Proof for Proposition 1]{Proof for Proposition \ref{prop:gmm}}\label{apdx:prop:gmm}

	Applying \eqref{eq:linear_substitution} to substitute the conditional density of $(Y_i(1),X_i)$ given $U_i$ and the conditional density of $Y_i(0)$ given $U_i$, the infeasible moment condition \eqref{eq:original_moment} becomes \begin{align*}
	0 &= \sum_{d=0,1} \int_{\mathbb{R}^3} m (y',y,d,x;\theta) \sum_{k=1}^K f_{Y(1),X|U}(y,x|u^k) \cdot f_{Y(0)|X}(y'|u^k) \cdot p_{D,U}(d,k) dy dy' dx \\
	&= \sum_{d=0,1} \sum_{k=1}^K \sum_{j=1}^K \sum_{j'=1}^K \frac{\tilde{\lambda}_0(j,k) \tilde{\lambda}_1(j,k)}{p_{D,Z}(0,j) p_{D,Z}(1,j')} \cdot p_{D,U}(d,k) \\
	& \hspace{15mm} \cdot \int_{\mathbb{R}^3} m (y',y,d,x;\theta) \cdot f_{Y,D,Z}(y,0,z^{j}) \cdot f_{Y,D,X,Z}(y',1,x,z^{j'})  dy dy' dx \\
	&= \E \left[ \tilde{m} \big( (Y_i, D_i, X_i, Z_i), (Y_{i'}, D_{i'}, X_{i'}, Z_{i'}) ; \theta, \tilde{\lambda}, p \big)\right]
	\end{align*}
	where \begin{align*}
	\tilde{m} \big( W_i, W_{i'} \big) &= \sum_{d=0,1} \sum_{k=1}^K p_{D,U}(d,k) \sum_{j=1}^K \sum_{j'=1}^K \frac{\tilde{\lambda}_0(j,k) \tilde{\lambda}_1(j,k)}{p_{D,Z}(0,j) p_{D,Z}(1,j')} \\
	& \quad \cdot \left( 0.5 \cdot m (Y_{i'},Y_i,d,X_{i'};\theta) \mathbf{1}\{D_i=0, Z_i \in \mathcal{Z}^j, D_{i'}=1, Z_{i'}\in \mathcal{Z}^{j'}\} \right. \\
	& \qquad \left. + 0.5 \cdot m (Y_i,Y_{i'},d,X_i;\theta) \mathbf{1}\{D_{i'}=0, Z_{i'} \in \mathcal{Z}^j, D_i=1, Z_i \in \mathcal{Z}^{j'}\} \right).
	\end{align*}
	\qed

	\subsection[Proof for Theorem 2]{Proof for Theorem \ref{thm:est}}\label{apdx:thm:est}

	With $M = M_Y \cdot M_X$, construct a $M \times M_Z$ matrix $\mathbf{H}_d$ that collects conditional probabilities $h(m,m'|d,j)$ across $(m,m')$ and $j$, for $d=0,1$: \begin{align*}
		\mathbf{H}_d = \begin{pmatrix} h(1,1|d,1) & \cdots & h(1,1|d,K) \\
		\vdots & \ddots & \vdots \\
		h(M_X,M_Y|d,1) & \cdots & h(M_X,M_Y|d,K) \end{pmatrix}.
	\end{align*}
	By Assumption \ref{ass:assignment}, $\mathbf{H}_d$ admits the decomposition $\mathbf{H}_0 = \Gamma_0 \cdot \Lambda_0$ and $\mathbf{H}_1 = \Gamma_1 \cdot \Lambda_1$, where \begin{align*}
	\Gamma_d = \begin{pmatrix} \gamma_{Y(d)}(1|1) \cdot \gamma_X(1|1) & \cdots & \gamma_{Y(d)}(1|K) \cdot \gamma_X(1|K) \\
	\vdots & \ddots & \vdots \\
	\gamma_{Y(d)}(M_Y|1) \cdot \gamma_X(M_X|1) & \cdots & \gamma_{Y(d)}(M_Y|K) \cdot \gamma_X(M_X|K) \end{pmatrix} \quad \text{for } d=0,1.
	\end{align*} 
	Let $\widehat{\Gamma}_0, \widehat{\Gamma}_1, \widehat{\Lambda}_0, \widehat{\Lambda}_1$ denote the matrix stackings built from $\hat{\gamma}_{Y(0)}(\cdot | \cdot), \dots, \hat{\lambda}_{1}(\cdot | \cdot)$ and $\mathbb{H}_0, \mathbb{H}_1$ denote the matrix stackings of $\hat{h} (\cdot | \cdot)$. Then, from the definition of the finite mixture estimator \eqref{eq:NMF}, \[
	\left\| \mathbb{H}_0 - \widehat{\Gamma}_0 \cdot \widehat{\Lambda}_0 \right\|_F + \left\| \mathbb{H}_1 - \widehat{\Gamma}_1 \cdot \widehat{\Lambda}_1 \right\|_F = \min_{\Gamma, \Lambda} \left\| \mathbb{H}_0 - \Gamma_0 \cdot \Lambda_0 \right\|_F + \left\| \mathbb{H}_1 - \Gamma_1 \cdot \Lambda_1 \right\|_F.
	\]
	
	The following shows that $\big\| \widehat{\Lambda}_0 - \Lambda_0 \big\|_F$ is $O_p \big( n^{-\frac{1}{2}} \big)$ up to some permutation of columns. We can apply the same argument to $\widehat{\Lambda}_1$. The proofs for the lemmas are provided in the Supplemental Appendix. 
	
	\begin{lemma}\label{lemma:est_H}
	Let Assumptions \ref{ass:assignment}-\ref{ass:condind}, \ref{ass:id}.a hold. Then, ${\left\| \Gamma_0 \Lambda_0 - \widehat{\Gamma}_0 \widehat{\Lambda}_0 \right\|_F}^2 = O_p \left(\frac{1}{\sqrt{n}} \right)$.
	\end{lemma}

	\begin{lemma}\label{lemma:est_rotation}
	Let Assumptions \ref{ass:assignment}-\ref{ass:id} hold. Then, $\big\| \widehat{\Gamma}_0 - \Gamma_0 A \big\|_F = O_p \left(\frac{1}{\sqrt{n}} \right)$ with some $K \times K$ matrix \begin{align}
	A = \begin{cases}
	\Lambda_0 \left( \widehat{\Lambda}_0\right)^{-1}, & \text{ if } {\Gamma_0}^\intercal \widehat{\Gamma}_0 \widehat{\Lambda}_0 \text{ is invertible} \\
	\mathbf{I}_K, & \text{ if } {\Gamma_0}^\intercal \widehat{\Gamma}_0 \widehat{\Lambda}_0 \text{ is not invertible}
	\end{cases} \label{eq:rotation}
	\end{align}
	with $\mathbf{I}_K$ being the $K \times K$ identity matrix.
	\end{lemma}

	\begin{lemma}\label{lemma:est_identityrotation}
	Let Assumptions \ref{ass:assignment}-\ref{ass:id} hold. Then, the $K \times K$ matrix $A$ defined in \eqref{eq:rotation} converges to a permutation matrix at the rate of $n^{-\frac{1}{2}}$, as $n \to \infty$. 
	\end{lemma}

	\noindent By rearranging the columns of $\widehat{\Gamma}_0$ and the rows of $\widehat{\Lambda}_0$ so that $A$ converges to $\mathbf{I}_K$,   \begin{align*}
		&\big\| \Lambda_0 - \widehat{\Lambda}_0 \big\|_F \\
		&\leq \Big\| \Lambda_0 - \left({\Gamma_0}^\intercal \Gamma_0 \right)^{-1} {\Gamma_0}^\intercal \widehat{\Gamma}_0 \widehat{\Lambda}_0 \Big\|_F + \Big\|  \left({\Gamma_0}^\intercal \Gamma_0 \right)^{-1} {\Gamma_0}^\intercal \widehat{\Gamma}_0 \widehat{\Lambda}_0 - \widehat{\Lambda}_0 \Big\|_F \\
		&\leq \Big\| \left({\Gamma_0}^\intercal \Gamma_0 \right)^{-1} {\Gamma_0}^\intercal \Big\|_F \cdot \big\| \Gamma_0 \Lambda_0 - \widehat{\Gamma}_0 \widehat{\Lambda}_0 \big\|_F \\
		&\hspace{4mm} + \big\| \widehat{\Lambda}_0 \big\|_F \cdot \left(\Big\| \left({\Gamma_0}^\intercal \Gamma_0 \right)^{-1} {\Gamma_0}^\intercal \left(\widehat{\Gamma}_0 - \Gamma_0 A \right) \Big\|_F + \Big\| \left({\Gamma_0}^\intercal \Gamma_0 \right)^{-1} {\Gamma_0}^\intercal \Gamma_0 \left( A - \mathbf{I}_K \right) \Big\|_F \right) \\
		&= \Bigg( \Big\| \left({\Gamma_0}^\intercal \Gamma_0 \right)^{-1} {\Gamma_0}^\intercal \Big\|_F + \big\| \widehat{\Lambda}_0 \big\|_F \cdot \Big\| \left({\Gamma_0}^\intercal \Gamma_0 \right)^{-1} {\Gamma_0}^\intercal \Big\|_F + \big\| \widehat{\Lambda}_0 \big\|_F \Bigg) \cdot O_p \left( \frac{1}{\sqrt{n}} \right).
	\end{align*}
	\qed

	\subsection[Proof for Theorem 3]{Proof for Theorem \ref{thm:clt}}\label{apdx:thm:clt}

	From Assumption \ref{ass:regular}.a-d, $\hat{\theta}$ is consistent for $\theta^0$. From consistency of $\big( \hat{\theta}, \hat{\tilde{\lambda}}, \hat{p}, \hat{\mu} \big)$, the orthogonality of the moment function $\psi$, and Assumption \ref{ass:regular}.b,e-f, \begin{align*}
		\sqrt{n} \left( \hat{\theta} - \theta^0 \right) &= - \left( \Psi^\intercal A \Psi \right)^{-1} \Psi^\intercal A \cdot \sqrt{n} \bar{\psi}_n(\theta^0, \tilde{\lambda}^0, p^0, \mu^0) + o_p(1).
	\end{align*}
	Applying the $U$-statistics asymptotic theory with Assumption \ref{ass:regular}.g, the asymptotic normality is established. The aymptotic variance is \[
	\Sigma_\theta = \left( \Psi^\intercal A \Psi \right)^{-1} \Psi^\intercal A \cdot \Sigma_\psi \cdot A^\intercal \Psi \left( \Psi^\intercal A \Psi \right)^{-1}
	\]
	where $\Sigma_\psi = 4 \Var \left( \tilde{\psi} \big(w;\theta, \tilde{\lambda}, p, \mu \big) \right)$. A plug-in estimator $
	\widehat{\Sigma}_\theta = \left( \widehat{\Psi}^\intercal A \widehat{\Psi} \right)^{-1} \widehat{\Psi}^\intercal A \cdot \widehat{\Sigma}_\psi \cdot A^\intercal \widehat{\Psi} \left( \widehat{\Psi}^\intercal A \widehat{\Psi} \right)^{-1}
	$
	with \begin{align*}
	\tilde{\psi}_{-i,n} (\theta) &= \frac{1}{n-1} \sum_{i' \neq i} \psi \big(W_i, W_{i'}; \theta, \hat{\tilde{\lambda}}, \hat{p}, \hat{\mu} \big), \\
	\widehat{\Sigma}_\psi &= \frac{4}{n} \sum_{i=1}^n \left( \tilde{\psi}_{-i,n} \big( \hat{\theta} \big) - \bar{\psi}_n \big( \hat{\theta} \big) \right) \cdot \left( \tilde{\psi}_{-i,n} \big( \hat{\theta} \big) - \bar{\psi}_n \big( \hat{\theta} \big) \right)^\intercal, \\
	\widehat{\Psi} &= \frac{1}{n} \sum_{i=1}^n \frac{\partial}{\partial \theta} \psi \big(W_i, W_{i'} ;\theta,\hat{\tilde{\lambda}}, \hat{p}, \hat{\mu} \big) \Big|_{\theta = \hat{\theta}} 
	\end{align*}
	is consistent.

	\qed					

	\bibliographystyle{plainnat}
	\bibliography{mylit}
\end{document}